\documentclass[]{emulateapj}
\usepackage{psfig,natbib,apjfonts}
\newcommand{\Msun}{\mbox{\,$\rm{M_{\odot}}$}} 
\newcommand{\Rsun}{\mbox{\,$\rm{R_{\odot}}$}} 
\newcommand{\Lsun}{\mbox{\,$\rm{L_{\odot}}$}} 
\newcommand{\Lbol}{\mbox{\,$\rm{L_{bol}}$}} 
\newcommand{\Lx}{\mbox{\,$\rm{L_{x}}$}} 

\newcommand{\Xsun}{\mbox{\,$\rm{X_{\odot}}$}}
\newcommand{\Tstar}{\mbox{\,$T_{*}$}}
\newcommand{\Teff}{\mbox{\,$T_{eff}$}}
\newcommand{\Rstar}{\mbox{\,$R_{*}$}} 
\newcommand{\Reff}{\mbox{\,$R_{2/3}$}} 
\newcommand{\Rtrans}{\mbox{\,$R_{t}$}} 

\newcommand{\Minit}{\mbox{\,$M_{init}$}} 
\newcommand{\Mdot}{\mbox{\,$\dot{M}$}}

\newcommand{\Mdotc}{\mbox{\,$\dot{M}_{cl}$}}
\newcommand{\Mdots}{\mbox{\,$\dot{M}_{sm}$}}

\newcommand{\vturb}{\mbox{\,$v_{turb}$}}

\newcommand{\vinf}{\mbox{\,v$_{\infty}$}}
\newcommand{\logg}{\mbox{\,$\log{g}$}}
\newcommand{\logN}{\mbox{\,$\log{N}$}}
\newcommand{\logL}{\mbox{\,$\log{L}$}}
\newcommand{\E}[1]{\mbox{\,$\rm x 10^{#1}$}}
\newcommand{\XH}{\mbox{\,$X_{H}$}}
\newcommand{\XHe}{\mbox{\,$X_{He}$}}
\newcommand{\XC}{\mbox{\,$X_{C}$}}
\newcommand{\XN}{\mbox{\,$X_{N}$}}
\newcommand{\XO}{\mbox{\,$X_{O}$}}

\newcommand{\XS}{\mbox{\,$X_{S}$}}
\newcommand{\XP}{\mbox{\,$X_{P}$}}
\newcommand{\XFe}{\mbox{\,$X_{Fe}$}}

\newcommand{\Htwo}{\mbox{\rm{H}$_2$}}
\newcommand{\HI}{\mbox{\rm{\ion{H}{1}}}}

\newcommand{\HeII}{\ion{He}{2}}

\newcommand{\CII}{\ion{C}{2}}
\newcommand{\CIII}{\ion{C}{3}}
\newcommand{\CIV}{\ion{C}{4}}
\newcommand{\NI}{\ion{N}{1}}

\newcommand{\NIV}{\ion{N}{4}}
\newcommand{\NV}{\ion{N}{5}}

\newcommand{\OI}{\ion{O}{1}}

\newcommand{\OIV}{\ion{O}{4}}
\newcommand{\OV}{\ion{O}{5}}
\newcommand{\OVI}{\ion{O}{6}}

\newcommand{\SiIV}{\ion{Si}{4}}

\newcommand{\PV}{\ion{P}{5}}
\newcommand{\SII}{\ion{S}{2}}

\newcommand{\SIV}{\ion{S}{4}}

\newcommand{\SVI}{\ion{S}{6}}

\newcommand{\FeIII}{\ion{Fe}{3}}

\newcommand{\FeVI}{\ion{Fe}{6}}

\newcommand{\FeVIII}{\ion{Fe}{8}}
\newcommand{\FeIX}{\ion{Fe}{9}}
\newcommand{\FeX}{\ion{Fe}{10}}

\newcommand{\eg}{\emph{e.g.}}
\newcommand{\ie}{\emph{i.e.}}
\newcommand{\doublet}{$\lambda\lambda$}
\newcommand{\singlet}{$\lambda$}
\newcommand{\res}{\mbox{\,$\Delta\lambda$}}

\newcommand{\EBMV}{\mbox{\,$E_{\rm{B-V}}$}}

\newcommand{\Rv}{\mbox{\,$R_v$}}

\newcommand{\tlusty}{TLUSTY}
\newcommand{\cmfgen}{CMFGEN}

\newcommand{\Hbeta}{H$\beta$}

\newcommand{\Lya}{Ly$\alpha$}
\newcommand{\Lyb}{Ly$\beta$}

\newcommand{\Zsun}{\mbox{\,$\rm{Z_{\odot}}$}}
\newcommand{\Telec}{\mbox{\,$T_{e}$}}
\newcommand{\nelec}{\mbox{\,$n_{e}$}}
\newcommand{\kms}{\mbox{\,$\rm{km\:s^{-1}}$}}

\newcommand{\Msunyr}{\mbox{\,$\rm{M_{\odot}\:yr^{-1}}$}}
\newcommand{\flam}{\mbox{\,$\rm{erg\:s^{-1}\:cm^{-2}\:\AA^{-1}}$}}

\newcommand{\svi}{\SVI\ \doublet 933,44}
\newcommand{\ciiia}{\CIII\ \singlet 977}
\newcommand{\ciiib}{\CIII\ \singlet 1175}
\newcommand{\ovi}{\OVI\ \doublet 1032,38}
\newcommand{\pv}{\PV\ \doublet 1118,28}
\newcommand{\nv}{\NV\ \doublet 1238,43}
\newcommand{\siiv}{\SiIV\ \doublet 1394,1402}
\newcommand{\oiv}{\OIV\ \doublet 1339,43}
\newcommand{\ov}{\OV\ \singlet 1371}
\newcommand{\civ}{\CIV\ \doublet 1548,51}
\newcommand{\lmco}{LMC-SMP~1}
\newcommand{\lmct}{LMC-SMP~29}
\newcommand{\lmcf}{LMC-SMP~50}
\newcommand{\lmcs}{LMC-SMP~76}
\newcommand{\lmca}{LMC-SMP~83}
\newcommand{\smco}{SMC-SMP~1}
\newcommand{\smce}{SMC-SMP~3}
\newcommand{\smcf}{SMC-SMP~5}
\newcommand{\smct}{SMC-SMP~22}

\begin{document}
\submitted{To be published in the Astrophysical Journal}

\title{CENTRAL STARS OF PLANETARY NEBULAE IN THE MAGELLANIC CLOUDS: A
  DETAILED SPECTROSCOPIC ANALYSIS\footnote{Based on observations made
  with the NASA-CNES-CSA Far Ultraviolet Spectroscopic Explorer and
  archival data. FUSE is operated for NASA by the Johns Hopkins
  University under NASA contract NAS5-32985.}}

\author{J.E. Herald, L. Bianchi}
\vspace{1mm}
\affil{Department of Physics and Astronomy, The Johns Hopkins University}
%\authoraddr{3400 N. Charles St., Baltimore, MD 21218-2411}

\begin{abstract}
We observed five central stars of planetary nebulae (CSPN) in the
Large Magellanic Cloud (LMC) and three in the Small Magellanic Cloud
(SMC) with the \emph{Far Ultraviolet Spectroscopic Explorer} (FUSE),
in the range 905---1187~\AA.  We performed a model-based analysis of
these spectra in conjunction with \emph{Hubble Space Telescope} (HST)
spectra in the UV and optical range to determine stellar and nebular
parameters.  The signature of hot ($T \gtrsim 2000$~K) circumstellar
molecular hydrogen is found in the FUSE spectra of most objects.  We
also find evidence of X-rays in the wind of \lmcs.
\end{abstract}

\keywords{stars: AGB and Post-AGB --- stars: atmospheres
  --- stars:  individual (\lmco, \lmct, \lmcf, \lmcs, \lmca, \smco,
  \smce, \smcf, \smct) ---
  ultraviolet: stars}

\section{INTRODUCTION}\label{sec:intro}

Central Stars of Planetary Nebulae (CSPN) residing in the Magellanic
Clouds (MCs) are an important asset to the study of CSPN evolution for
two main reasons: their distances are well constrained, allowing their
physical parameters (\eg, radius and luminosity) to be determined with
much less uncertainty than their Galactic counterparts, and the
metallicities of their host galaxies are lower than in the
Milky Way.  Metallicity is expected to influence a star's evolution by
setting the efficiency of radiative driving during the phases when
the star has a stellar wind: the asymptotic giant branch (AGB) phase,
the post-AGB phase, and the [Wolf-Rayet] phase ([WR]).  Higher
metallicity, and thus wind radiative acceleration, should increase the
star's mass-loss rate and wind velocity, altering the layers exposed
and chemical yields as well as the kinematics of the nebular shell.  This has
implications for galaxy evolution, through chemical enrichment and the
dynamic interactions between the star's ejected material and the
surrounding interstellar medium (ISM).  The large fraction of stars
that pass through these phases make their contribution to the galactic
chemical evolution significant (see, \eg, \citealp{marigo:01} for a
discussion).

The far-UV wavelength range is useful for studying hot PN
nuclei, as they emit the bulk of their observable flux in this range,
and it contains many strong lines which are diagnostics of stellar
parameters.  It is also uncontaminated by nebular continuum emission \citep{bianchi:97,herald:04a},
which affects wavelengths longer than \Lya, allowing us to
characterize the physical parameters of both the nebular and central
star components through detailed quantitative modeling.

Motivated by the above considerations, we (\citealp{herald:04a} ---
hereafter, HB04) performed a spectroscopic analysis on far-UV and UV
data of seven CSPN in the Large Magellanic Cloud (LMC) using data
taken with the \emph{Far-Ultraviolet Spectroscopic Explorer} (FUSE).
As a continuation of that work, we have obtained FUSE observations of
five additional LMC CSPN (Bianchi's D034 program) and three Small
Magellanic Cloud (SMC) CSPN (Bianchi's C056 program).  These targets
were selected from those expected to be the brightest (in the far-UV)
from the photoionization studies of
\citet{dopita:91a,dopita:91b,dopita:97,vassiliadis:96s,vassiliadis:98s}.
The additional LMC targets (D034) were selected to
expand the temperature coverage of the HB04 LMC sample to higher
temperatures, and with the SMC sample we gain objects in a much lower
metallicity environment.  Together with the LMC sample of Bianchi's
FUSE proposal B001, analyzed by HB04, these programs cover all the
known CSPN in the MC observable by FUSE.  Our analysis is similar to
that of HB04, and the reader is referred to that work for more details
and background information.

This paper is arranged as follows.  The observations and data
reduction are described in \S~\ref{sec:obs}.  A comparison of the
spectra of the objects is presented in \S~\ref{sec:description}.  Our
models and parameter determinations are described in
\S~\ref{sec:modeling}.  Individual object results are presented in
\S~\ref{sec:results}.  The implications of our results
are discussed in \S~\ref{sec:discussion} and our conclusions in
\S~\ref{sec:conclusions}.

\section{OBSERVATIONS AND REDUCTION}\label{sec:obs}

For this work, we used far-UV spectra taken with FUSE and archive UV and
optical spectra taken with the \emph{Hubble Space Telescope's} (HST)
\emph{Faint Object Spectrograph} (FOS) or \emph{Space Telescope
Imaging Spectrograph} (STIS), which we now describe.

\subsection{Far-UV Spectra}\label{sec:fuvobs}

The FUSE observations (programs C056 and D034) of our sample stars are
summarized in Table~\ref{tab:fuseobs}.  They are some of the dimmest
stellar objects yet observed by FUSE, and necessitated long
integration times.  FUSE covers the wavelength range 905---1187~\AA\
at a spectral resolution of $R \approx 20,000$ ($\sim$ 15 \kms). The
instrument is described by \citet{moos:00} and its on-orbit
performance is discussed by \citet{sahnow:00}. FUSE collects light
concurrently in four different channels (LiF1, LiF2, SiC1, and SiC2),
each of which is divided into two segments (A \& B) recorded by two
detectors, covering different subsets of the above range with some
overlap.  We used FUSE's LWRS (30\arcsec$\times$30\arcsec) aperture.
These data, taken in ``time-tag'' mode, have been calibrated using the
FUSE data reduction pipeline, efficiency curves and
wavelength solutions (CALFUSE v2.4).

The spectra of these targets suffer from significant contamination
from terrestrial airglow lines (such as \OI\ and \NI), and scattered
solar emission features (\ovi\ and \ciiib), which often interfere with
the spectroscopic analysis by obscuring the stellar features.  For the
LMC sample, we use ``night-only'' data (\ie, taken when the
spacecraft was on the dark side of the Earth), which minimize this
contamination, although some residual airglow lines (typically \OI)
can still appear (an example of the difference between day and night
observations can be found in HB04).

Because of their faintness, most of our targets required many
orbit-long exposures, each of which typically had low count-rates and
thus signal-to-noise ratios.  Each calibrated extracted sequence was
checked for unacceptable count-rate variations (a sign of detector
drift), incorrect extraction windows, and other anomalies.  Segments
with problems were not included in further steps.  
The default FUSE pipeline attempts to model the background measured
off-target and subtracts it from the target spectrum. In some
cases, the background is taken to be the model scattered-light scaled
by the exposure time.

All the ``good'' calibrated segments/exposures were combined using FUSE
pipeline routine and overlapping ranges of the different
channels were then compared for consistency.  There are several
regions of overlap between the different channels, and these were used
to ensure that the continuum matched. For the fainter targets, often
the continuum levels for different detectors did not agree.  The
procedure we typically adopted was to trust the LiF1a absolute flux
calibration (which is the most reliable --- \citealp{sahnow:00}), and
scale the flux of the other detectors if needed.  The LiF1b segment is
sometimes affected by an artifact known as ``the worm'' (FUSE Data Handbook
v1.3) and was omitted if the effect was found to be severe.
The SiC1 detector seemed frequently off-target, and other segments
offered higher S/N data in the same range, so its data was often not
used.  In the end, we typically used the SiC2 data for wavelengths
below $\sim1000$~\AA.  Longwards of 1080~\AA\ LiF2a data was employed.
For the intermediate region (\ie, 1000--1080~\AA), data from LiF1b,
LiF2b, or both were used (data from one segment was omitted if it was
discrepant with that of the SiC detectors).  The region between 1083
and 1087~\AA\ is not covered by the LiF detectors, and as the SiC
detectors in this range were off-target, we have omitted this region
(appearing as gaps in the plots).  Table~\ref{tab:fuseobs}
shows which segments and portions were used for the combined spectrum.
Data from near the detector edges were also omitted if they looked
inconsistent.  

Some individual objects warrant more discussion. The SMC observations
had numerous difficulties.  These observations occurred during a
period of time after FUSE lost one of its reaction wheels, affecting
its pointing, but before jitter correction had been introduced into
the FUSE calibration pipeline to account for such effects.  For \smco,
it appears only the LiF1 was aligned with the central star with any
reliability (the fine error sensor camera, used for guiding, is part
of the LiF1 channel, so the location of the source is always aligned
in the LiF1 apertures).  Other detectors seemed to have gathered
nebular emission lines with little or no stellar continuum for this
object.  We therefore only present the LiF1 data for this object but
note that the LiF1b (longer wavelength) segment is often affected by
``the worm'' (see above).  Although the worm is not obviously present
in the LiF1b observations of this object, the low count rate prevents a
definitive statement regarding this.  During the observation of \smct,
there was significant target drift as evidenced by inconstant count
rates, and the LiF2 and SiC2 detectors seem to have been off-target.
It also appears that another star was present in the aperture during
much of the observation (discussed in \S~\ref{sec:results_smc}). We
present here data from the SiC1 and LiF1 detectors, but we stress that
the absolute fluxes are uncertain due to what appears to be the target
drifting in and out of the aperture throughout the observation.  For
\smcf, no useful data was acquired.  In the case of \lmct, voltage and
alignment problems resulted in no reliable FUSE data being collected.

The FUSE spectra presented in this paper were obtained by combining the
good data from different segments, weighted by detector sensitivity, and
rebinning to a uniform dispersion of 0.05~\AA.  

\subsection{UV and Optical Data}\label{sec:uvobs}

Archival data from the HST's \emph{Faint Object Spectrograph} (FOS)
and \emph{Space Telescope Imaging Spectrograph} (STIS) instruments
were also used longwards of 1200~\AA, as summarized in
Table~\ref{tab:uvobs}.  The FOS archive data were taken in the
1\arcsec\ aperture, whose actual diameter following the COSTAR
installation is 0.86\arcsec.  The angular sizes of the LMC PN are
typically $\lesssim$1\arcsec, so the FOS spectroscopy includes the
entire nebula (the exception is \lmca, see Table~\ref{tab:neb}).  The
utilized FOS dispersers include G130H ($\res\sim1$~\AA), G190H
($\res\sim1.5$~\AA), G270H ($\res\sim2$\~\AA) and G400H
($\res\sim3$~\AA).  For \lmca, medium-resolution STIS data is
available, taken through the 52\arcsec x0.2\arcsec\ aperture with the
G140L ($\res\sim1.3$~\AA), G230L ($\res\sim1.5$~\AA), G430L
($\res\sim1.6$~\AA) and G750L $\res\sim1.6$~\AA) gratings.

Generally, the flux levels of the FUSE and FOS/STIS data in the region
of overlap are in good agreement for the LMC observations, except for
\lmca, which is known to be variable
(\S~\ref{sec:results}).  The SMC sample contains more discrepancies,
however.  For \smct, the UV flux level is much higher than that of the
far-UV (discussed more in \S~\ref{sec:results_smc}).  For \smce\ and \smco,
the FUSE flux levels are higher by about a factor of 2 and 1.5
(respectively) than the FOS data.  \emph{International Ultraviolet
Explorer} (IUE) spectra of these objects display similar flux levels
as those of FOS, therefore we we have scaled the FUSE flux levels to
that of the FOS in these instances.

\section{Description of the Spectra}\label{sec:description}

\begin{figure}
\begin{center}
\epsscale{1}
%\begin{turn}{90}
\rotatebox{0}{
\plotone{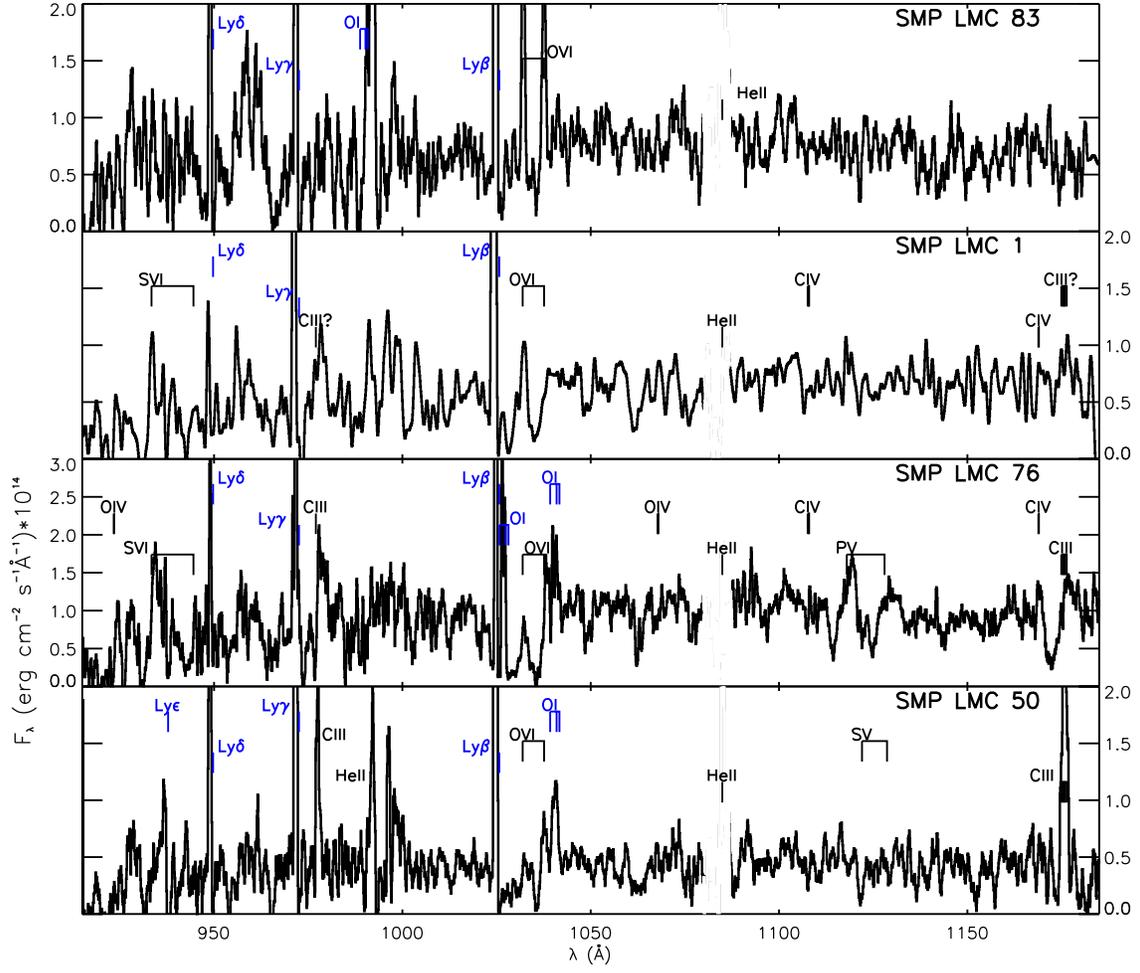}}
%\end{turn}
\caption{FUSE spectra of our LMC sample, velocity shifted to the rest frame of
  the central star (Table~\ref{tab:neb}), and re-binned to a
  resolution of 0.10\AA.  The more prominent stellar features and nebular
  emission lines are marked by black labels, airglow features are marked by
  the blue labels.  Most objects seem to display
  wind lines to some extent.  Nebular \OVI\ emission appears in the spectra of
  \lmca.}\label{fig:lmc_fuse}
\end{center}
\end{figure}

\begin{figure}
\begin{center}
\epsscale{1}
%\begin{turn}{90}
\rotatebox{0}{
\plotone{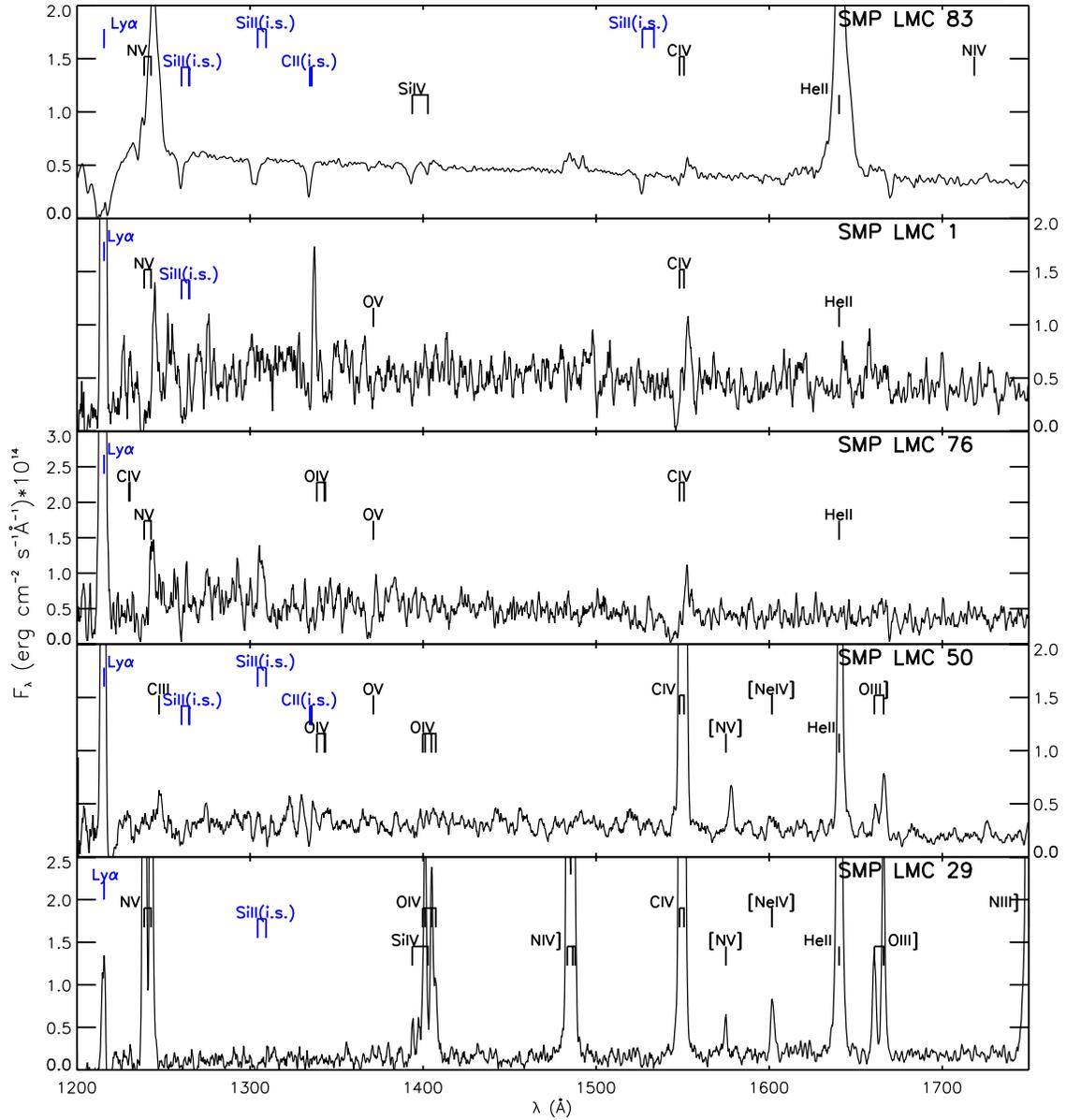}}
%\end{turn}
\caption{FOS and STIS (for \lmca) spectra  of our LMC sample, velocity shifted
  to the rest frame of the central star (Table~\ref{tab:neb}), and re-binned to a
  resolution of 0.25\AA.  The
  more prominent stellar features and nebular emission lines are
  marked with black labels, interstellar absorption features are
  marked with blue/gray labels.  Most objects display wind lines to
  some extent.  \CIV\ \doublet 1548-51 appears as a P-Cygni profile in
  most.  The spectra of \lmcf\ and \lmct\ show emission lines from
  nebulae in high states of ionization.  }\label{fig:lmc_uv}
\end{center}
\end{figure}

The reduced FUSE spectra for the LMC sample, along with line
identifications, are shown in Fig.~\ref{fig:lmc_fuse}, and the
corresponding UV spectra are shown in Fig.~\ref{fig:lmc_uv}.  

Wind features are definitely seen in the far-UV spectra of most
stars: \svi, \ciiia, and \svi\
are visible in \lmco\ and \lmcs.  The latter also shows \ciiib\ and
\pv.  For \lmcf, the only feature obviously
attributable to a stellar wind is \ovi.  \lmca\ shows also nebular
emission in \OVI, implying a very hot central star (temperatures of $\Teff \gtrsim
125$~kK are needed to generate these lines through photoionization ---
\citealp{chu:04}).  In the far-UV spectra, \Htwo\ absorption makes the
continuum level very difficult to set (\S~\ref{sec:htwo}).

In the UV range, \lmcf\ shows some nebular features
and an underlying continuum, but strong stellar features are
absent.  \lmca, \lmco, and \lmcs\ all display a P-Cygni profile of \nv, as well
as \civ\ of different strengths.  The blue edges of the
P-Cygni profiles in both the far-UV and UV spectra indicate the
terminal wind velocities of all objects are $\vinf \simeq
500-1000$~\kms, except in the case of \lmca\ ($\vinf \simeq
1600$~\kms) and \smce\ ($\vinf \ge 2000$~\kms).

\begin{figure}
\begin{center}
\epsscale{1}
%\begin{turn}{90}
\rotatebox{0}{
\plotone{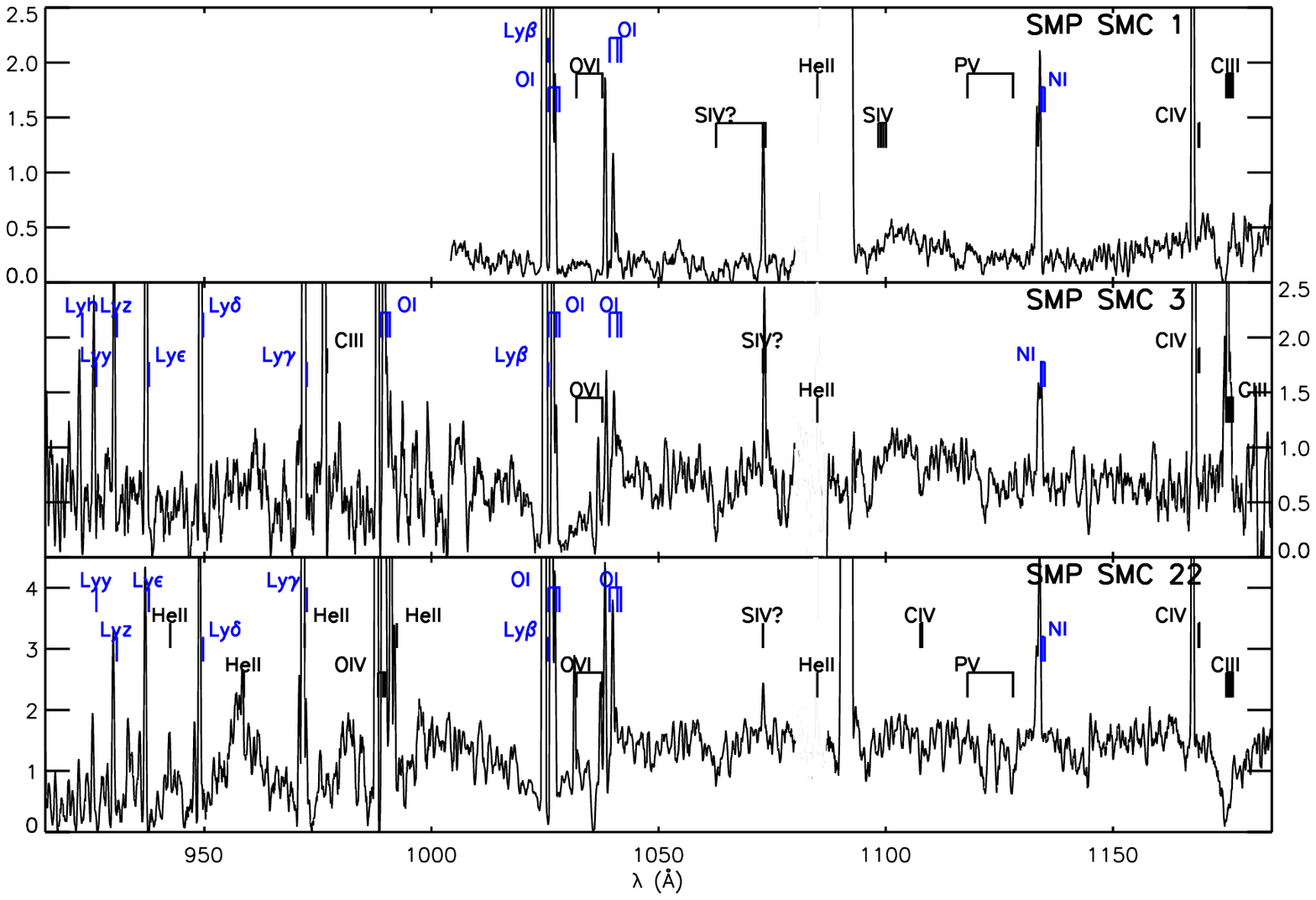}}
%\end{turn}
\caption{FUSE spectra of our SMC sample, velocity shifted to the rest frame of
  the central star (Table~\ref{tab:neb}), and re-binned to a
  resolution of 0.10\AA.  The more prominent stellar features and nebular
  emission lines are marked by black labels, airglow features are marked by
  the blue/gray labels. Note the \OVI\ \doublet 1032-38 nebular
  emission features of \smct.  }\label{fig:smc_fuse}
\end{center}
\end{figure}

\begin{figure}
\begin{center}
\epsscale{1}
%\begin{turn}{90}
\rotatebox{0}{
\plotone{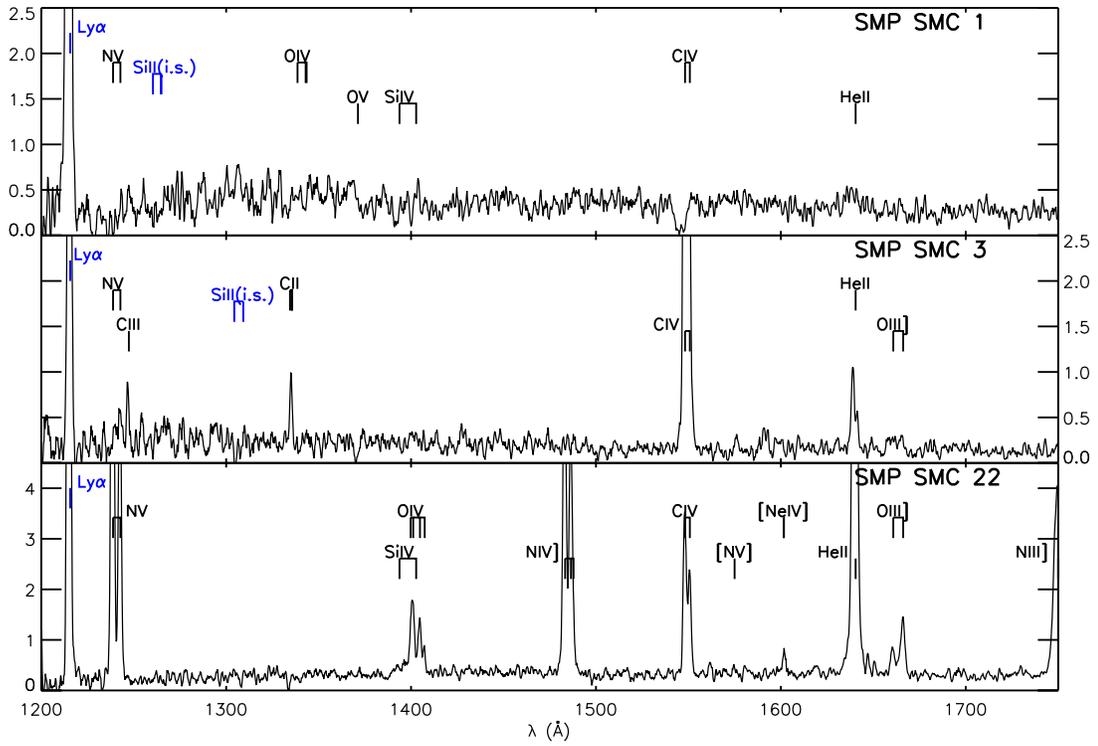}}
%\end{turn}
\caption{UV spectra (FOS) of our SMC sample, velocity shifted to
  the rest frame of the central star (Table~\ref{tab:neb}), and re-binned to a
  resolution of 0.25\AA.
  The more prominent stellar features and
  nebular emission lines are marked with black labels, airglow lines or
  interstellar absorption features are marked with blue/gray labels.  
  }\label{fig:smc_uv}
\end{center}
\end{figure}

The reduced FUSE, and archive UV, spectra for the SMC sample are shown
in Fig.~\ref{fig:smc_fuse} and Fig.~\ref{fig:smc_uv}, respectively.
In the FUSE observations, nebular and airglow features contaminate the
spectra (especially at short wavelengths).  \ciiib\ is present
in absorption in the spectra of \smco\ and \smct.  \ovi\ nebular lines
are also seen in the latter.  The only wind features in the UV spectra
are \civ\ seen in \smco.  \smct\ shows a more highly-ionized nebular
spectrum than that of \lmca.

\section{MODELING}\label{sec:modeling}

Modeling the central stars of the MC CSPN presents some challenges.
First, their nebulae are typically compact ($\lesssim 1$\arcsec), and
in most cases entirely contained in the FOS aperture used 
in the archival data available.  Their UV and optical spectra are
frequently contaminated by the nebular continuum and lines, which
obscure and sometimes entirely mask the stellar spectrum in these
regions.  Thus, one must rely on a smaller set of stellar features in
the far-UV and UV to determine the stellar parameters.  Second, the
far-UV region, while not influenced by nebular continuum emission, is
affected by absorption by electronic transitions of sight-line
molecular hydrogen (\Htwo).  In our objects, such absorptions cover
the entire far-UV region.  The analysis of the spectra consists
of modeling the central star spectrum, the nebular continuum emission
longwards of 1200~\AA, and the sight-line Hydrogen (atomic and
molecular) shortwards concurrently and to find a consistent solution.
However, we discuss each in turn for clarity.  The reddening has to be
derived concurrently, but it is small in all cases, and therefore it does
not contribute much to the uncertainties.

Throughout this paper, we adopt a LMC distance of $D=50.6$~kpc
\citep{feast:91}, an SMC distance of $D=60.0$~kpc \citep{harries:03},
an LMC metallicity of $z = 0.4\Zsun$ \citep{dufour:84}, and an SMC
metallicity of $z = 0.1\Zsun$, with values of ``solar'' abundances
from \citet{grevesse:98}.

As with HB04, we assume Galactic foreground extinctions of 0.05
\citep{bessell:91}, with the remainder being from the Magellanic
Cloud.  We use the \citet{fitzpatrick:99} reddening law for the LMC
and the \citet{cardelli:89} reddening law for the Galaxy ($\Rv=3.1$)
extinction, extrapolated down to far-UV wavelengths.

\subsection{Molecular and Atomic Hydrogen}\label{sec:htwo}

Absorption due to atomic (\HI) and molecular hydrogen (\Htwo) along
the sight-line often complicates the far-UV spectra of these objects.
Toward a CSPN, this sight-line material typically consists of
interstellar and circumstellar components (\eg,
\citealp{herald:02,herald:04a,herald:04b}).  Material comprising the
circumstellar \HI\ and \Htwo\ presumably was ejected from the star
earlier in its history, and is thus important from an evolutionary
perspective.

The \Htwo\ transitions in the FUSE range are from the Lyman ($B^1
\Sigma^+_u$--$X^1 \Sigma^+_g$) and Werner ($C^1 \Pi^{\pm}_u$--$X^1
\Sigma^+_g$) series.  We (HB04) found that, for our previous LMC CSPN, a
significant amount of \emph{hot} (\ie, $T\gtrsim2000$~K) \Htwo\ lies
along the sight-lines, presumably associated with the nebula.  At
such temperatures, higher vibrational states become populated and the
absorption pattern is much more complex, obscuring the entire far-UV
continuum spectrum (see Fig.~5 of HB04).  With care, the effects of
such a gas can be accounted for and strong stellar features can be
discerned, but weaker stellar features will be unrecoverable.

We have modeled the \Htwo\ toward our sample CSPN in the following
manner.  For a given column density ($N$) and gas temperature ($T$),
the absorption profile of each line is calculated by multiplying the
line core optical depth ($\tau_0$) by the Voigt profile $[H(a,x)]$
(normalized to unity) where $x$ is the frequency in Doppler units and
$a$ is the ratio of the line damping constant to the Doppler width
(the ``b'' parameter).  The observed flux is then $F_{obs} =
\exp{[-\tau_{0}H(a,x)]} \times F_{intrinsic}$.  We first assume the
presence of an interstellar component with $T=80$~K (corresponding to
the mean temperature of the ISM --- \citealp{hughes:71}) and \vturb =
10\kms.  The column density of this interstellar component is
estimated by fitting its strongest transitions.  If additional
absorption features due to higher-energy \Htwo\ transitions are
observed, a second (hotter) component is added, presumed to be
associated with the nebula.  We thus velocity-shift the \Htwo\
absorption features of the circumstellar component to the radial
velocity (from Table~\ref{tab:neb}) of the particular star.
HB04 found that, typically, the absorption pattern of the
circumstellar hot \Htwo\ was adequately fit by absorption models with
$\Teff \simeq 2000$~K and $\log{N} \simeq 16.7$~cm$^{-2}$, and the
reader is referred to that work for a more detailed discussion of the
\Htwo\ fitting.  We note that our terminology of ``circumstellar'' and
``interstellar'' components is a simplification, indicating that we
assume the ``cool'' component to be mostly interstellar and the
``hot'' component mostly circumstellar.  However, the column density
derived for the cooler ``interstellar'' component may also include
circumstellar \Htwo.  Our derived parameters for sight-line Hydrogen
are shown in Table~\ref{tab:hydrogen}.   Note for some
objects (\eg, \lmct\ and \smco), we have not attempted to fit the \Htwo\
spectrum, due to the lack of useful FUSE data.  If a number appears in
the table without uncertainties, that value has been assumed rather
than derived.  Molecular hydrogen
disassociates around $T \simeq 2500$, so such high temperatures are
signs of non-equilibrium conditions (note our models assume thermal
populations).  Such \Htwo\ characteristics are observed in shocked
regions.  
A shocked region is likely complex, containing areas of \Htwo\ of
differing characteristics.  Because of the high complexity of the
absorption pattern at these hot temperatures, we did not attempt to
fit every feature, but the absorption pattern in the far-UV as a
whole, to a point where the stellar features could be identified and
modeled.

When possible, \HI\ column densities are determined from the \Lya\ and
\Lyb\ features, which encompass both the interstellar and possible nebular
components.  The absorption profiles of \HI\ are calculated in
a similar fashion as described above for \Htwo. 

We also list in Table~\ref{tab:hydrogen} the reddenings implied by our
measured column densities of \HI\ using the relationship $\left<
N(\HI)/\EBMV \right> = 4.8\E{21}$~cm$^{-2}$~mag$^{-1}$
\citep{bohlin:78}, which represents typical conditions in the
(Galactic) ISM.

\subsection{The Nebular Continuum}\label{sec:nebcont}

Nebular parameters taken from the literature for the program objects 
are compiled in Table~\ref{tab:neb}.  They include the angular size of
the nebula $\theta$, the nebular radius $r_{neb}$, the expansion
velocity $v_{exp}$, the electron density $n_{e}$, the electron
temperature $T_e$, the \Hbeta\ flux $F_{H\beta}$, the Helium to
Hydrogen ratio and the doubly to singly ionized Helium ratio.  Values
in \textbf{bold} are our derived values (see below).  We also list the
reddenings determined from literature values of the
logarithmic extinction at H$\beta$ using the relation $c_{H\beta}
= 1.475 \EBMV$.

In some cases, the nebular continuum significantly contributes to the
observed spectra at wavelengths $\gtrsim 1200$~\AA\ and must be
estimated to fit the data (\S~\ref{sec:cs}).  

\begin{figure}
\begin{center}
\epsscale{1}
%\begin{turn}{90}
\rotatebox{0}{
\plotone{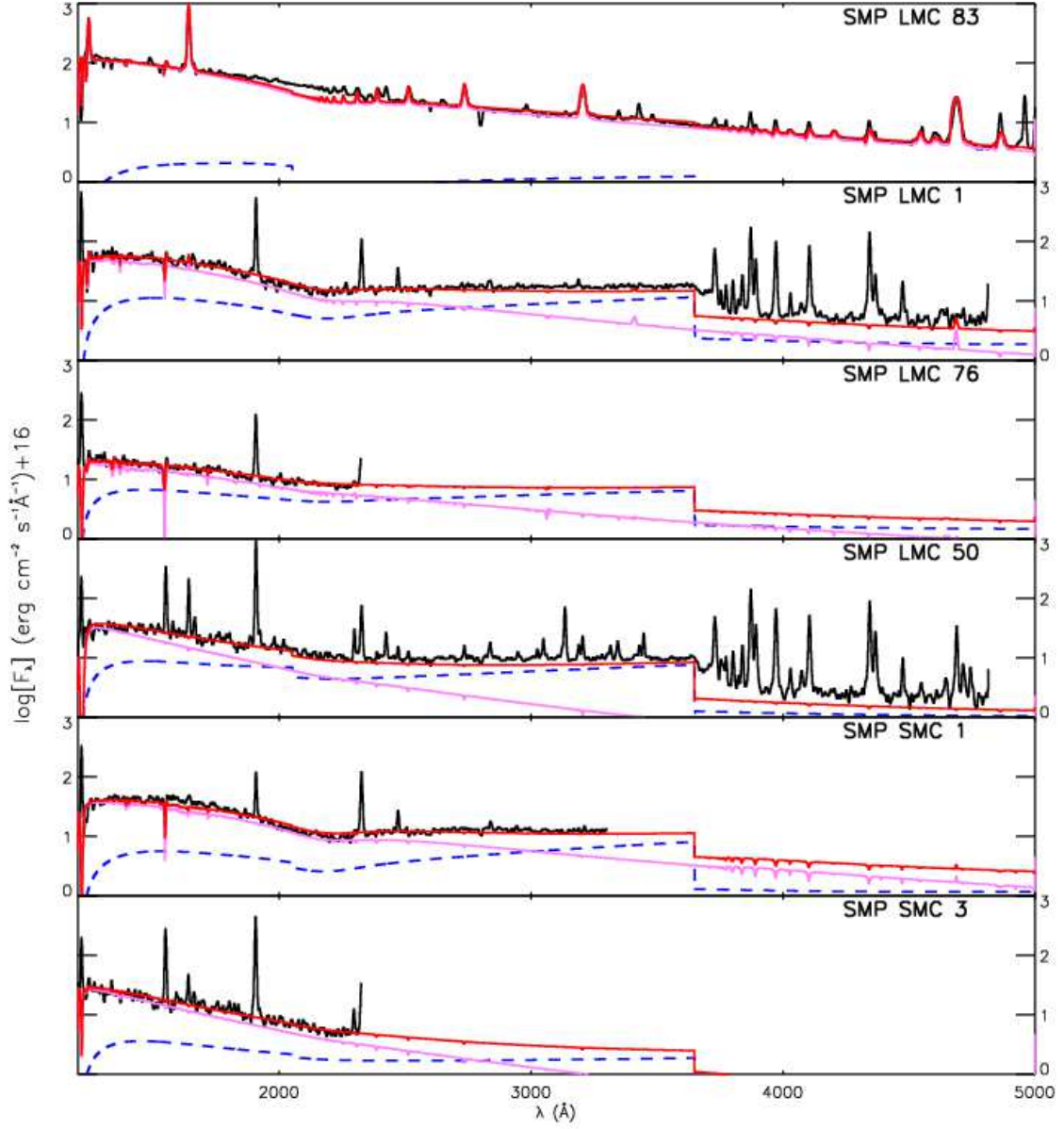}}
%\end{turn}
\caption{The MC CSPN spectra (black) are shown along with our stellar
  (pink/grey) and nebular continuum models (blue dashed).  The sum
  of the models is shown by the red line.  All models have been
  reddened with our determined values for \EBMV\
  (Table~\ref{tab:mod_param}), and the effects of hydrogen absorption
  (Table~\ref{tab:hydrogen}) have been applied.  The spectra have been
  convolved with a 5~\AA\ Gaussian for clarity.  Note the logarithmic
  flux scale.  }\label{fig:neb}
\end{center}
\end{figure}

The nebular continuum emission has been modeled using the code
described in \citet{bianchi:97}, which accounts for two-photon, H and
He recombination, and free-free emission processes.  The computed
emissivity coefficient of the nebular gas is scaled as an initial
approximation to the total flux at the Earth by deriving the emitting
volume from the dereddened absolute \Hbeta\ flux.  In the case of
\lmca, the nebula was not entirely contained within the aperture used
for the STIS observations, and so we scaled the continuum by an
appropriate geometrical factor ($4A/\pi\theta^{2}$ where $A$ is the
area of the aperture in square arcseconds).  $T_e$ and \nelec\ from
the literature are used as initial inputs, and adjusted if necessary.
\lmco\ and \lmcs\ have no published values of the He$^{2+}$/He$^{+}$
ratio.  For these, we found adopting a ratio of 0 to produce
satisfactory fits (these are relativity cooler objects).  The values
of $F_{H\beta}$ and c$_{H\beta}$ are perhaps the most uncertain, and
we finally scale the nebular flux with the assumption that it is
responsible for the majority of the flux at longer wavelengths.  The
nebular continuum flux level is further constrained not to exceed the
P-Cygni troughs in the UV range.  Our combined stellar and nebular
models, along with the observations, are shown in Fig.~\ref{fig:neb}.

\subsection{The Central Stars}\label{sec:cs}

To model the spectra of the central stars, we used CMFGEN
\citep{hillier:98,hillier:99b,hillier:03}, a line-blanketed code
suitable for an extended, spherically-symmetric expanding atmosphere.
We also used the hydrostatic structures from \tlusty\
\citep{hubeny:95} models as input to the CMFGEN models.  \tlusty\
calculates the atmospheric structure assuming a plane-parallel
geometry.  The detailed workings of the code are explained in the
references above, here we give a brief descriptions of the salient
features.

For \cmfgen\ models, the fundamental photospheric/wind parameters
include \Teff, \Rstar, the mass loss rate \Mdot, the elemental
abundances, the velocity law and the wind terminal velocity, \vinf.
\Rstar\ is taken to be the inner boundary of the model atmosphere
(corresponding to a Rosseland optical depth of $\sim20$).  The
\emph{stellar temperature} $\Tstar$ is related to the luminosity and
radius by $L = 4\pi\Rstar^2\sigma\Tstar^4$, and the \emph{effective
temperature} (\Teff) is similarly defined but at a radius
corresponding to a Rosseland optical depth of 2/3 (\Reff).  The
luminosity is conserved at all depths, so $L =
4\pi\Reff^2\sigma\Teff^4$.

We assumed what is essentially a standard velocity law for the stellar
wind $v(r) =
\vinf(1-r_0/r)^\beta$ where $r_0$ is roughly equal to \Rstar.  The
choice of velocity law mainly affects the profile shape, not the total
optical depth of the wind lines, and does not greatly influence the
derived stellar parameters.  Once a velocity law is specified, the
density structure of the wind $\rho(r)$ is then parameterized by the
mass-loss rate \Mdot\ through the equation of continuity:
$\Mdot=4\pi\Rstar^2 \rho(r)v(r)$.  However, one actually can only
derive \Mdots=(\Mdotc/$\sqrt{f}$) from the models, where \Mdots\ and
\Mdotc\ are the smooth and clumped mass-loss rates, and the degree of
clumpiness is parametrized by the \emph{filling factor} ($f$).  Both
theory \citep{owocki:88,owocki:94} and observations indicate that
radiation driven winds are clumped to some extent.  Unless otherwise
noted, \Mdot\ refers to \Mdots\ throughout this paper.

It has been found that wind models with the same \emph{transformed
radius} \Rtrans\ [$\propto \Rstar(\vinf/\Mdot)^{2/3}$]
\citep{schmutz:89} and \vinf\ have the same ionization structure,
temperature stratification and spectra (aside from appropriate
scalings with \Rstar\ --- \citealp{schmutz:89,hamann:93}).  Thus,
once the velocity law and abundances are set, one parameter may be
fixed (say \Rstar) and parameter space can then be explored by varying
only the other two parameters (\eg, \Mdot\ and \Teff).  For opacities
which are proportional to the square of the density, the optical depth
of the wind scales as $\propto \Rtrans^{-3}$, so \Rtrans\ can be
thought of as an optical depth parameter.

In the CMFGEN models presented by HB04, a constant scale height ($h$)
was adopted, which connects the spherically extended hydrostatic outer
layers to the wind.  The gravity is related to the scale height as $h
\propto g^{-1}$.  Our models here differ in this respect, as instead
of adopting a constant scale height, we have
used the hydrostatic structures as generated from our \tlusty\ models
in the CMFGEN models, in the manner as described by \citet{bouret:03}.
Because of the severe \Htwo\ absorption in the far-UV, and the masking
of the stellar flux by the nebular continuum at longer wavelengths,
the gravity cannot be well constrained in our analysis.  Thus, we
adopt gravities typical of CSPNe of the given \Teff, guided by the
\logg-\Teff\ relations of \citet{vassiliadis:94} (hereafter, VW94)
(\eg, $\logg=4.7$ for $\Teff \simeq 55$~kK, $\logg=6.0$ for $\Teff
\simeq 100$~kK, etc.).  The wind features (used as diagnostics)
are not very sensitive to the gravity, and typically the
observed wind features can be fit with models of the same \Teff\ but
with gravities differing by over a magnitude.  Line profiles in the
optical, which would provide good diagnostics of the gravity, are not
available for our sample (the G750L STIS data of \lmca\ does cover the
optical range, but due to its emission line spectrum, provides no
useful gravity diagnostics).

\subsubsection{Abundances and Model Ions}\label{sec:abundances}

We note here that throughout this work, the nomenclature $X_i$
represents the mass fraction of element $i$, and `\Xsun'' denotes the
solar value.  Following HB04, we adopted two model abundance patterns
for the MC CSPN.  The first, appropriate for ``H-rich'' CSPN, have
solar abundances for H and He but assume LMC/SMC metallicities of $z =
0.4/0.1$~\Zsun\ for the metals.  For the second, corresponding to
``H-deficient'' (or H-poor) CSPN, we have assumed no hydrogen and an
abundance pattern of \XHe/\XC/\XN/\XO = 0.55/0.37/0.01/0.08 with
LMC/SMC metallicities of $z = 0.4/0.1$~\Zsun\ for the other elements
(\eg, S, Si, \& Fe).  The abundances for the H-rich and H-deficient
grids are summarized in Table~\ref{tab:abund}.  H-deficient abundances
(which are enriched in oxygen and carbon) are signs of He-burning.
Normal, ``H-rich'' abundances can be seen when the star is either H or
He-burning.  

We mention here that there are no diagnostics in the far-UV or UV
spectra of our objects that allow us to make a statement regarding
their hydrogen/helium ratio.  The handful of wind lines upon which we are
basing our parameter determinations do not allow us to make firm
statements regarding the specific abundances of a given element, but
in most cases we are able to determine which grid is more appropriate
based on the strength of carbon and oxygen features.

For the model ions, both \tlusty\ and \cmfgen\ utilize the concept of
``superlevels'', whereby levels of similar energies are grouped
together and treated as a single level in the rate equations (after
\citealp{anderson:89}).  Ions and the number of levels and superlevels
included in the model calculations, as well references to the atomic
data, are given in the Appendix (\S~\ref{sec:atomic}).

\subsubsection{Diagnostics}\label{sec:diag}

Our sample covers a wide range of temperatures and mass-loss
rates.  As different diagnostics are used in different
parameter regimes, we shall discuss the diagnostics used in a
case-by-case basis (\S~\ref{sec:results}) and here we discuss only the
more general diagnostics.

For stars showing strong wind features in their spectra, the terminal wind
velocity (\vinf) can be estimated from the blue edge of the P-Cygni
absorption features.  Strong P-Cygni profiles unobscured by \Htwo\
absorption or nebular emission are rare in the far-UV spectra of our
sample.  We mostly used \CIV\ \doublet 1548-51 (when present) for our
initial estimate of \vinf, and adjusted this value to match the wind
lines.  Due to lack of suitable diagnostics, we did not attempt
to rigorously constrain the degree of clumping in the winds of our
sample.  In the parameter regime of our objects, the model spectra
are mainly insensitive to $f$, except for \ovi\ in
the hotter models ($\Teff \gtrsim 50$~kK) and \PV\ \doublet 1118,28
for cooler models (HB04).  We note here that the smooth mass-loss rate
\Mdots\ is an upper limit of the actual mass-loss rate.

For the other parameters, we adopted the following method.  We
compared the observed spectra with the theoretical spectra of our
model grids to determine the stellar parameters \Teff\ and \Rtrans\
(in the case of \cmfgen\ models) or \Teff\ and \logg\ (in the cast of
\tlusty\ models).  Far-UV and UV flux levels (corrected for the
nebular contribution) were used to set \Rstar\
(since the distance is known), and thus \Mdot\ and $L$ can be
derived from \Rtrans.  Uncertainties in the quoted parameters reflect the range
of acceptable model fits.

Typically, one determines \Teff\ by the ionization of the CNO
elements.  For these objects, we typically use \CIII/\CIV\ and/or
\OIV/\OV/\OVI\ to constrain \Teff.  It is desirable to use diagnostics
from many elements to ensure consistency.  The presence or absence of
some features offer further constraints.  For example, for mass-loss
rates in the regime of our stars, \SiIV\ \doublet 1394,1402 disappears
at temperatures $\Teff \gtrsim 50$~kK, \SIV\ \doublet 1062,1075 at
$\sim$45~kK, and \PV\ \doublet\ 1118,28 at $\sim$60~kK.  We generally
use all wind lines to constrain \Mdot.

\section{STELLAR MODELING RESULTS}\label{sec:results}

Our derived central star parameters are summarized in
Table~\ref{tab:mod_param}.  We also indicate
the model abundances used for each object, and list the reddenings
determined by fitting the multiple components to the far-UV/UV
spectra. We now discuss the modeling diagnostics and results for the
individual objects. 

\subsection{Results for LMC CSPN}\label{sec:results_lmc}

\begin{figure}[htbp]
\begin{center}
\epsscale{1.0}
%\begin{turn}{90}
\rotatebox{0}{
\plotone{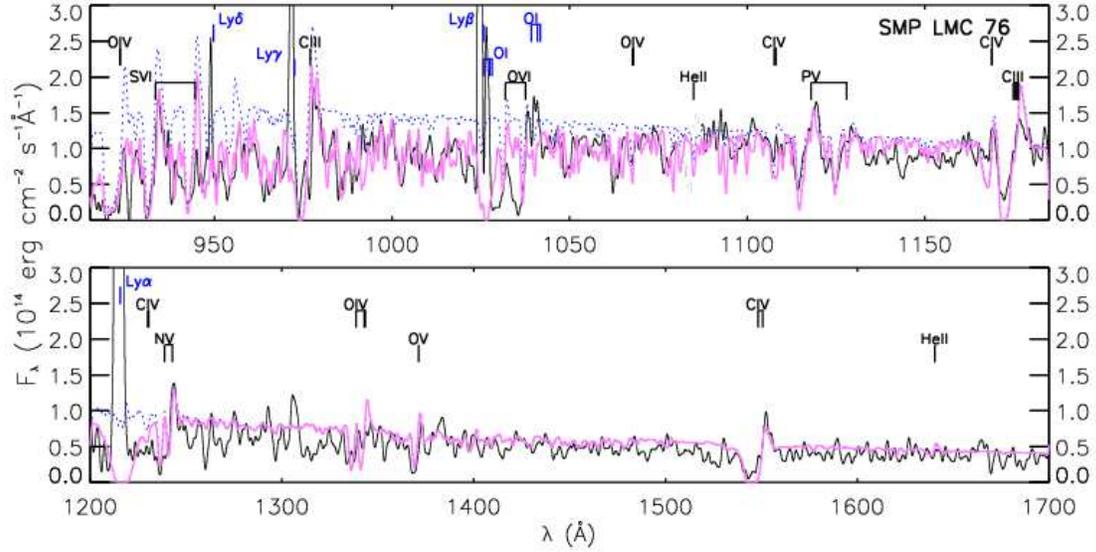}}
%\end{turn}
\caption{\lmcs:  The observations are shown (black) along with
  our stellar model, with and without our hydrogen absorption models
  applied (pink/gray and blue dotted, respectively).  The FUSE
  spectra have been convolved with a 0.6~\AA\ Gaussian for clarity.
}
\label{fig:lmc76}
\end{center}
\end{figure}

\noindent \emph{\lmcs\ (Fig.~\ref{fig:lmc76})}

\noindent The far-UV and UV spectra of this object are rich in wind
features: \SVI\ \doublet 933,44, \CIII\ \singlet 977 \& 1175, \PV\
\doublet 1118,28, \OVI\ \doublet 1032,38, \NV\ \doublet 1238,43, \OV\
\singlet 1371, and \CIV\ \doublet 1548,51.  The \OVI\ doublet has
particularly strong absorption (but some of the emission seen
includes \OI\ airglow lines).

The strong carbon and oxygen spectrum exhibited by this object are
best fit by models from our H-deficient grid (implying He-burning).
In the model shown ($\Teff = 50$~kK, $\Mdot=3.15\E{-8}$~\Msunyr) most
of the features mentioned above are fit adequately, except \pv\ (which
is weak) and \ciiib\ (which is strong).  Adjusting the
temperature/mass-loss rate within the quoted uncertainties alleviate
these problems, at the expense of degrading the fits of some of the
other diagnostics.  In cooler models the
\CIII\ features becomes too strong, unobserved \SiIV\ \doublet
1394,1402 appears, and \NV\ \doublet\ 1238-43 disappears.  Hotter
models ($\Teff \gtrsim 60$~kK) lose the \CIII\ lines.  Models fail to
duplicate the strong \OVI\ line in any case, which requires $\Teff
\gtrsim 70$~kK before it begins to be fit adequately.  This problem
is alleviated by including X-rays in the model atmosphere (discussed below).
Based on the \CIII\ and \CIV\ wind lines, we derive a
terminal wind velocity of $\vinf = 1250 \pm 250$~\kms.

Our parameters for this star fall between the ($M_i,M_f$) = (1.0,
0.578)~\Msun\ and (1.5, 0.626) (He-burning) tracks ($z=0.008$) of
VW94.  This, combined with a better fit of the observation with models
from our H-deficient (\ie, C and O-enriched) grid, cause us to
conclude \lmcs\ to be a He-burner.  This is in contrast to
\citet{dopita:97}, who concluded this object to be H-burning from
comparison with evolutionary tracks based on their higher derived
luminosity of $\log{L} = 3.735$~\Lsun.  They also inferred a very high
nebular carbon abundance for this object, which is consistent with our
findings that the wind is carbon-enriched.  \citet{bianchi:97}
determined $\Teff = 52$~kK, $\vinf = 1500$~\kms\ for this object by
modeling its UV and optical spectra, basically confirmed by our analysis.

The FUSE spectra of this star show the signature of hot molecular
hydrogen, which severely attenuates the stellar flux.  We have fit the
absorption pattern with circumstellar \Htwo\ component having a
temperature of $T\simeq 3000$~K.

\begin{figure}[htbp]
\begin{center}
\epsscale{.55}
\rotatebox{270}{
\plotone{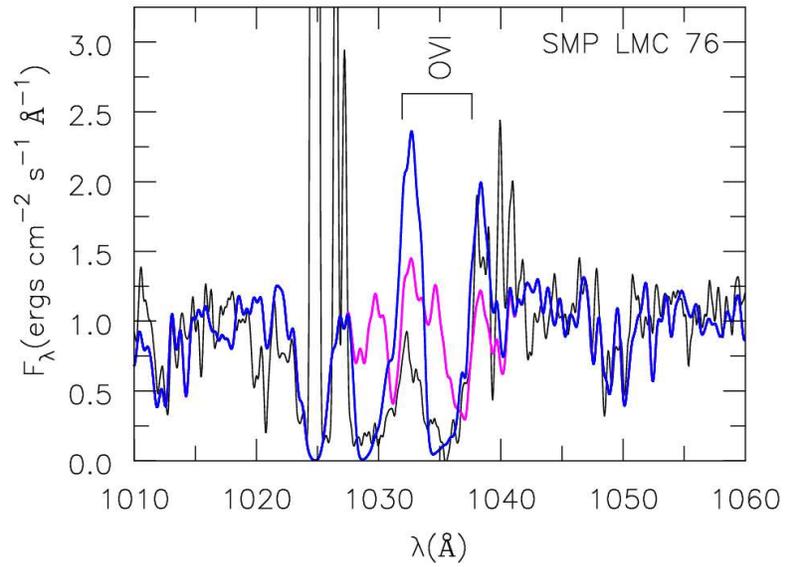}}
\caption{\lmcs:  The observations are shown (black) along with
  two models, one without X-rays (pink/gray) and one with X-rays of
  strength $\log{\Lx/\Lbol} \lesssim -7.0$ (blue/dark grey).  Without X-rays,
  the model fails to duplicate the strong P-Cygni absorption trough
  caused by \OVI\ \doublet 1032,38.  Including X-rays in the model
  calculations strengthens the \OVI\ feature while not affecting other
  diagnostics.
}
\label{fig:lmc76_xray}
\end{center}
\end{figure}

As mentioned above, our models failed to fit the \ovi\ line
simultaneously with the other, lower-ionization diagnostics.  It is
known that X-rays, arising from shocks believed to develop from
instabilities in the radiationally-driven wind, are needed to explain
observed \OVI\ line profiles in O and early B-stars
\citep{bianchi:02,garcia:04,bianchi:06iau2}.  We thus calculated some
models which include X-rays in the wind.  CMFGEN implements X-rays as
described in \citet{martins:05b}.  In summary, shock sources are
assumed to be distributed throughout the atmosphere once the wind
achieves a certain velocity, with emissivities taken from
\citet{raymond:77}. The shock parameters include the shock temperature
(which controls the energy distribution of the shocks), the volume
filling factor (which sets the level of emission), and the velocity of
the wind where the shocks are switched on.  For the starting velocity,
we have used $\sim400$~\kms, as radiative shocks are believed to
become important when the wind achieves this velocity.  We adopted a
shock temperature of 300~kK, typical of high energy photons in O type
stars (\citealp{schulz:03,cohen:03}), and adjusted the volume filling
factor until adequate fits of the \OVI\ feature were achieved.  CMFGEN
outputs the total X-ray luminosity, and the X-ray luminosities in the
energy range above 1 and 0.1~keV, for the fluxes both absorbed by the
wind, the X-ray flux which escapes.  For our adopted parameters, the
X-ray luminosity above 0.1~keV is about 5 times that above 1~keV.  We
found that for $\log{\Lx/\Lbol} \gtrsim -9.0$, the strength of the
absorption and the red emission components of the \OVI\ P-Cygni profile were
adequately duplicated (see example Fig.~\ref{fig:lmc76_xray}),
although the blue emission component is too strong (most probably due
to our simple velocity law).  For greater
\Lx, the emission feature was a bit stronger than observed.  Other
spectral features did not change until $\log{\Lx/\Lbol} \simeq -6.0$,
at which point \NIV\ \singlet 955 decreased in strength.  The \OVI\
feature did not change with respect to our model without X-rays for
$\log{\Lx/\Lbol} \lesssim -10.0$.  Note that these X-ray luminosities
represent the escaped X-ray flux, \ie, what would be observed. The
actual X-ray energy input into the wind is $\sim 500$ times more.
\citet{guerrero:01} discuss the possibility that shocks in the stellar
wind may explain the observed, point-source X-ray fluxes
($\log{\Lx/\Lbol} \lesssim -7.0$) of the central star of the Cat's Eye
nebula (NGC~6453).

\begin{figure}[htbp]
\begin{center}
\epsscale{1.0}
%\begin{turn}{90}
\rotatebox{0}{
\plotone{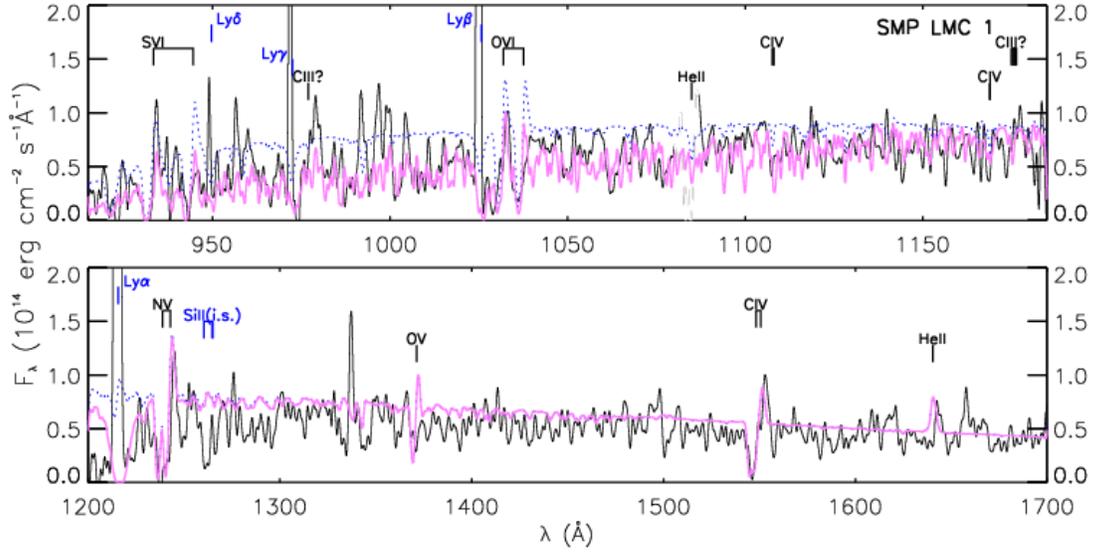}}
%\end{turn}
\caption{\lmco: Description follows that of Fig.~\ref{fig:lmc76}.
}
\label{fig:lmc1}
\end{center}
\end{figure}

\noindent \emph{\lmco\ (Fig.~\ref{fig:lmc1})}

\noindent The spectrum of \lmco\ is similar to that of \lmcs, but with
the wind features having a lower terminal velocity ($\vinf \lesssim
800$~\kms) and weaker oxygen signature.  We can achieve adequate fits
using either models from the H-rich or H-poor grid, with the former
requiring higher mass-loss rates.  We list both models in
Table~\ref{tab:mod_param}, but only show the H-poor model in the
figures.  The higher temperature limit ($\Teff=70$~kK) fits the \ovi\
absorption trough better.  For the lower temperatures ($\Teff =
55$~kK) and mass-loss rates, the \ciiib\ feature is too strong
compared with the observations.  In the UV, the \nv\ and \civ\ are
adequately fit with models in the quoted parameter range, but the \ov\
feature is always too strong (HB04 had similar problems with this line
with some of their sample) in the case of either abundance pattern.
The \oiv\ feature is similarly too strong in the H-poor model spectra,
while the carbon features are generally better-fit than with our
H-rich model grid (probably indicating our adopted oxygen abundance is
too high).  The FUSE spectrum shows the absorption pattern from hot
($T \simeq 3000$~K) molecular hydrogen, which severely depresses the
stellar flux.  In comparison with \lmcs, this object has a higher
luminosity and effective temperature, but a lower \vinf\ (mass-loss
rates are comparable).  This could be explained by our finding that
the wind of \lmcs\ seems more obviously chemically enriched, and thus
a higher opacity leading to a more efficient transfer of
radiation-to-wind momentum.

\begin{figure}[htbp]
\begin{center}
\epsscale{1.0}
%\begin{turn}{90}
\rotatebox{0}{
\plotone{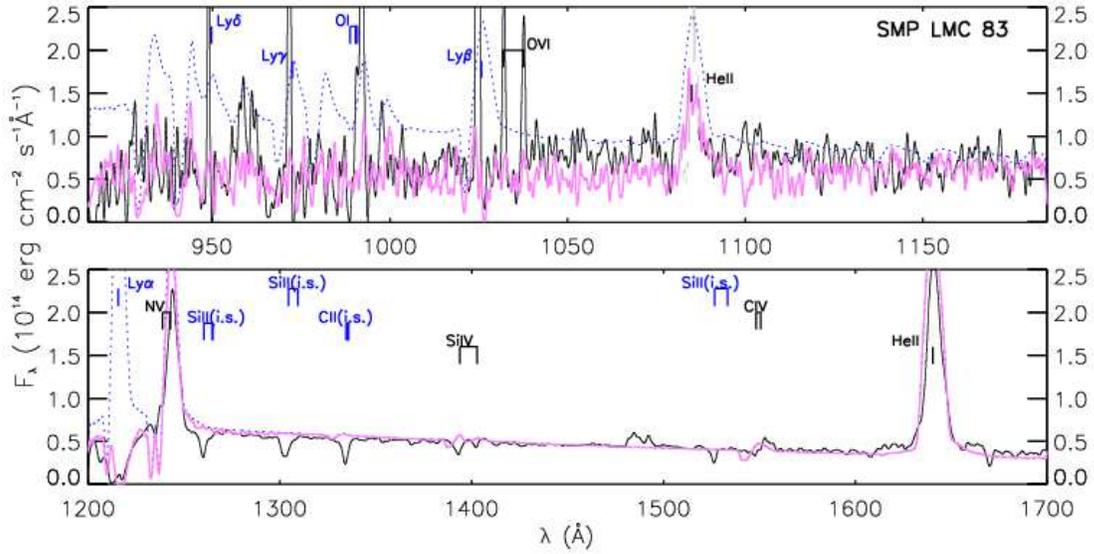}}
%\end{turn}
\caption{\lmca: Description follows that of Fig.~\ref{fig:lmc76}.
}
\label{fig:lmc83}
\end{center}
\end{figure}

\noindent \emph{\lmca\ (Fig.~\ref{fig:lmc83}):}

\noindent \lmca\ is a bit of an oddball, for which the classification
as a CSPN is a matter of current debate.  The object is notable for
multiple reasons.  Between 1993 and 1994, its luminosity increased
from $\log{L}=4.6$ to 5.6 (\Lsun).  The quiescent luminosity is
already extreme for a CSPN (most LMC CSPN have $\log{L} \lesssim
3.9$~\Lsun, \eg, \citealp{dopita:91a,dopita:91b}), but is less
luminous than any known massive WN (\eg, \citealp{hamann:98a}).
Furthermore, it is unique in that it displays a [WN] spectrum (other
LMC CSPN which have WR-type spectra show [WC] spectra).  It has been
the subject of a long-term monitoring campaign and UV spectra exist
spanning many epochs.  \citet{hamann:03} (hereafter, HPGR) performed
detailed UV and optical spectral modeling across these epochs, and
found the following parameters to be constant: $\Tstar = 112\pm20$~kK,
$\XH = 0.2$, $\XN = 3\E{-3}$, $\XC \leq 1\E{-4}$, $\XFe = 2\E{-4}$.
The increase in luminosity during outburst was apparently caused by an
increase in mass-loss rate, which increased the effective radius.  The
nature of \lmca\ is debatable, and HPGR discuss many alternatives to
the CSPN classification (\eg, single massive star, low-mass binary
system with mass transfer, double degenerate merger).  The main
argument against the CSPN classification is its surface nitrogen-rich
abundances, which are difficult to explain in the CSPN context.

\begin{figure}[htbp]
\begin{center}
\epsscale{0.25}
%\begin{turn}{90}
\rotatebox{270}{
\plotone{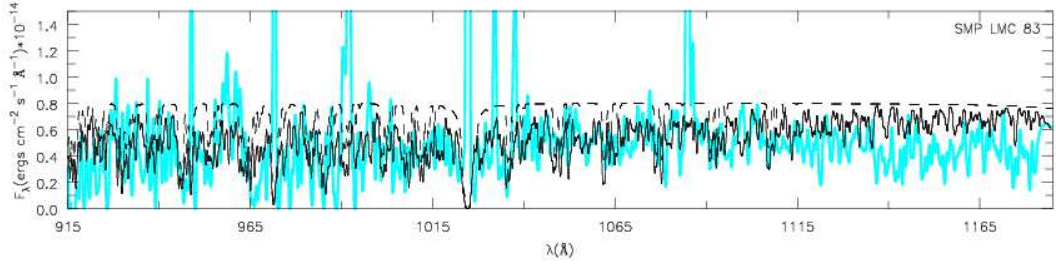}}
%\end{turn}
\caption{\lmca: A portion of the FUSE spectrum is shown (black)
  along with a flat model continuum, both with only an interstellar
  hydrogen absorption component applied (dashed) and with an
  additional hot circumstellar \Htwo\ component applied (green).  }
\label{fig:lmc83h2}
\end{center}
\end{figure}

In Fig.~\ref{fig:lmc83h2}, we show the absorption pattern of the
interstellar hydrogen absorption component, and an additional hot
circumstellar component applied to a flat continuum.  There does
appear to be an overall signature of hot \Htwo\ in the data, although
many minor features are not well fit.  The \Htwo\ absorptions are
significant, and no obvious strong stellar features are apparent in
the FUSE observations.  Because the FUSE range offers no further
stellar diagnostics, we have not performed much further modeling of this
object beyond that of HPGR. 

The FUSE flux levels are a factor of about 1.5 higher
compared to the STIS data.  \lmca\ was observed by STIS in 1999 and
again in 2000, and comparison of these two data sets shows the 1999
continuum level in the 1250-1550~\AA\ region to be about 95\% that of
the 2000 level measured 13 months later (the FUSE data was gathered
about 1.5 yr after the 2000 STIS observations).  As noted by
\citet{hamann:03}, there is very little change in the actual spectral
features within this time.  For this work we have used the 1999 STIS
data set, because its signal to noise is superior to that of the 2000
data set, and scaled the FUSE data to the STIS flux levels.

We first computed a model with parameters similar to what HPGR found
for the most recent epoch (2000): $\vinf=1600$~\kms, $\Rstar =
0.5$~\Rsun, $\Teff = 95$~kK, $\log{\Mdot} = -5.67$~\Msun, and the
constant parameters quoted above (abundances and \Tstar).  However,
rather than using their reddening (a Galactic foreground reddening of
$\EBMV = 0.03$~mag and an LMC reddening of $\EBMV = 0.13$~mag), we
used a smaller amount of reddening, to produce better agreement over
our expanded wavelength range.  The use of this less severe reddening
results in a slightly smaller derived radius (and corresponding
mass-loss rate) than those of HPGR.  This model produced several
\HeII\ features in the FUV which do not appear to be present in the
observations.  However, it is possible that severe absorption from a
hot circumstellar \Htwo\ gas (\S~\ref{sec:htwo}) could attenuate the
\HeII\ features to agree with the observations, as shown in
Fig.~\ref{fig:lmc83}.  Another possibility is that the mass-loss rate
has since declined to the point these features no longer appear in
emission.  Note that because we have essentially adopted the stellar
parameters of this object from HPGR and not performed any further stellar
modeling, we have omitted uncertainties for its stellar parameters in 
Table~\ref{tab:mod_param}.

\begin{figure}[htbp]
\begin{center}
\epsscale{1.0}
%\begin{turn}{90}
\rotatebox{0}{
\plotone{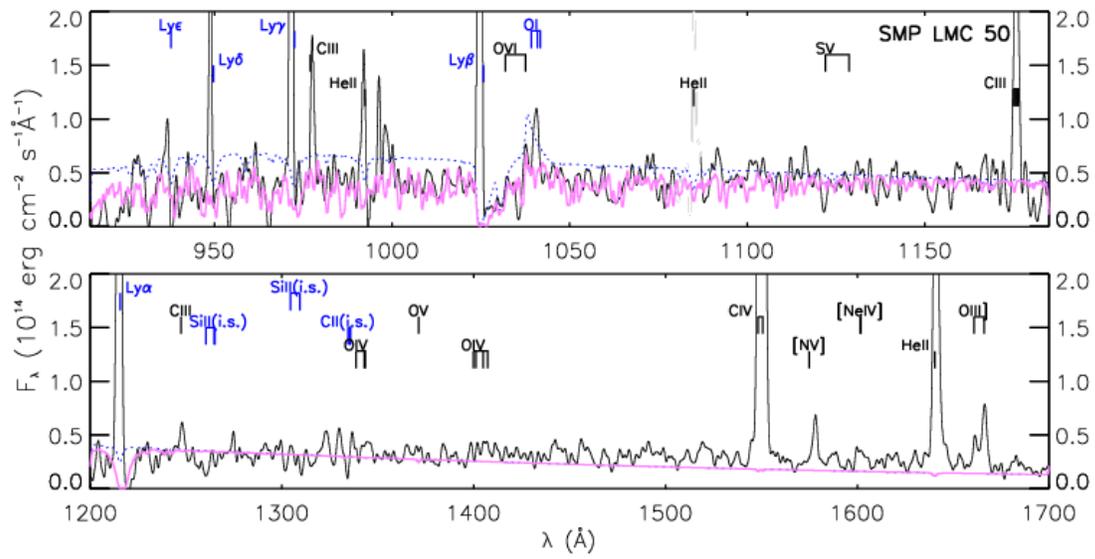}}
%\end{turn}
\caption{\lmcf:  Description follows that of Fig.~\ref{fig:lmc76}.
}
\label{fig:lmc50}
\end{center}
\end{figure}

\noindent \emph{\lmcf\ (Fig.~\ref{fig:lmc50})}

The photoionization temperature for this star is 100~kK (see
Table~\ref{tab:shock}), and we find that a H-rich model with $\Teff
\simeq 110$~kK, $\logg\simeq 6.0$, $\log{\Mdot} = -8.2$~\Msun,
$\vinf\simeq500$~\kms, $\Rstar\simeq 0.21$~\Rsun does an adequate job
at fitting the most apparent wind feature displayed --- that of the
\OVI\ doublet (see Fig.~\ref{fig:lmc_fuse}).  These parameters lie
slightly below the $\Minit = 2.0$~\Msun, $M = 0.67$~\Msun\ track of
VW94.  The rest of the theoretical spectrum is relatively featureless,
however, whereas the FUSE and FOS display some complexity.  Many of
the absorption features in the shorter wavelength region of the FUSE
spectrum can be attributed to molecular hydrogen, as shown in the
figure.  The poor quality of the observations make firm identification
of stellar and interstellar features difficult.  For example, the FOS
spectra show a P-Cygni-like feature at \singlet 1335, which might be a
\CII\ feature of stellar origin (indicating a much cooler object,
possibly a companion) or possibly a circumstellar absorption feature
in a relatively noisy spectrum that mimics an actual P-Cygni line.
The lack of strong, clean stellar features prohibit a more rigorous
determination of stellar parameters.

\noindent \emph{\lmct}

Along with \smct, the FOS spectra show show high-ionization lines
(\eg, \nv) and have extreme photoionization temperatures ($T \simeq
200$~kK).  Unfortunately, problems during the FUSE observation
(\S~\ref{sec:obs}) resulted in unreliable FUSE data for this object,
which we have not attempted to model.  We have also made no attempt to
fit the interstellar \Lya\ absorption feature.

\subsection{Results for SMC CSPN}\label{sec:results_smc}

\begin{figure}[htbp]
\begin{center}
\epsscale{1.0}
%\begin{turn}{90}
\rotatebox{0}{
\plotone{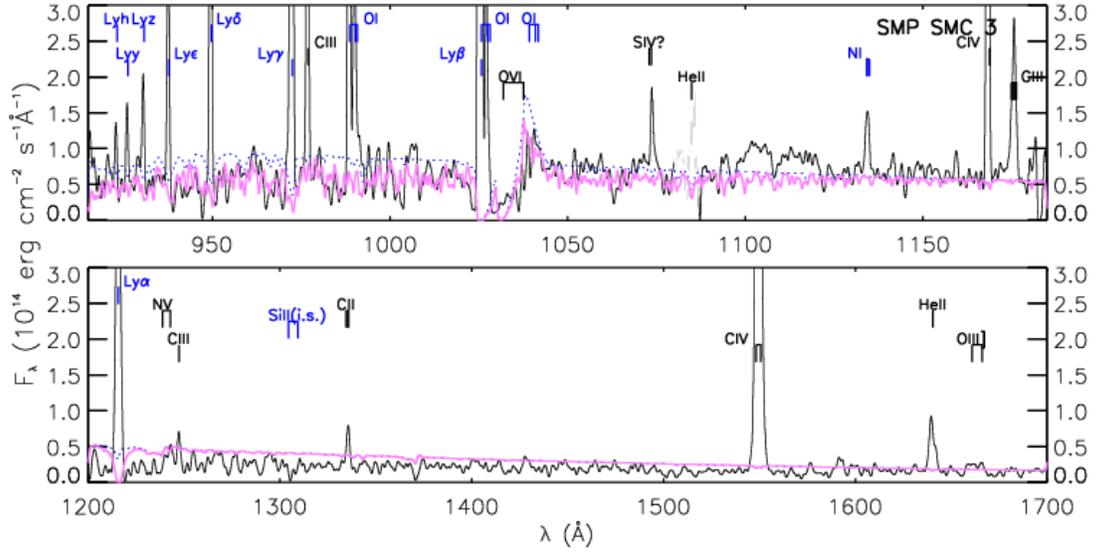}}
%\end{turn}
\caption{\smct: Description follows that of Fig.~\ref{fig:lmc76}.
}
\label{fig:smc3}
\end{center}
\end{figure}

\noindent \emph{\smce\ (Fig.~\ref{fig:smc3})}

There are no obvious wind features in the FOS spectrum of this object,
and the FUSE spectrum only obviously shows the \OVI\ resonance line.
The FUSE spectrum is also contaminated by several strong airglow
features, especially at shorter ($\lesssim 1000$~\AA) wavelengths, and
it shows a moderate \Htwo\ signature, which we have fitted.
This star has a relatively high
photoionization temperature of $\Teff = 92~$~kK
(Table~\ref{tab:shock}).  Given the paucity of wind diagnostics, we
can only speak in general terms about its parameters.  Models with
$90\leq \Teff \leq 100$~kK with $-8.0 < \log{\Mdot} <-7.0$~\Msun\ and
$\vinf \gtrsim 2000$~\kms\ adequately reproduce the \OVI\ feature.
Cooler temperatures and higher mass-loss rates typically show
unobserved spectral features, and lower mass-loss rates result in a
\OVI\ weak compared to observations.  Lower \vinf\ results in both
components of the \OVI\ doublet being resolved --- contrary to
observations.  However, the P-Cygni edge of the \OVI\ feature is
obscured by absorption due to \Lyb, so we cannot make a more precise
measurement of \vinf.  We conclude by saying a H-rich model with
temperatures consistent with the photo-ionization temperature with 
\Mdot\ and \vinf\ within the previously stated limits fit the FUSE and
FOS spectra adequately, with the caveat that these parameters are
constrained mainly by the \OVI\ doublet and the absence of other
strong features.  For comparison with evolutionary models, a $\Teff =
95$~kK model scaled to the observed FOS flux levels would have
$\Rstar \simeq 0.32$~\Rsun, $\logL \simeq 3.8$~\Lsun.  Comparison with the
$Z=0.004$, $\Minit = 1.5$~\Msun, $M=0.640$~\Msun\ track of VW94 
implies a post-AGB age of a few kyr.

For this object, as well as \lmcf, we have relied mainly on the \OVI\
doublet as a diagnostic.  In our discussion of \lmcs, we showed this
line could be heavily influenced by X-rays, which might add additional
uncertainty to parameters derived using this line.  However,
photoionization models indicate that \smce\ (and \lmcf) are
considerably hotter than \lmcs\ ($\Teff \gtrsim 90$ vs. $\sim50$~kK).
The absence of low-ionization spectral features supports the idea that
these two objects lie in this high temperature regime.  Preliminary
models computed by us indicate that, in this regime, the high
temperatures alone are adequate to reproduce the observed \OVI\
profile, and including reasonable amounts of X-rays in the model
atmospheres does not alter this profile significantly.  We plan a more
detailed investigation of the role of X-rays in the winds of CSPN to
be the subject of a future paper.

\begin{figure}[htbp]
\begin{center}
\epsscale{1.0}
%\begin{turn}{90}
\rotatebox{0}{
\plotone{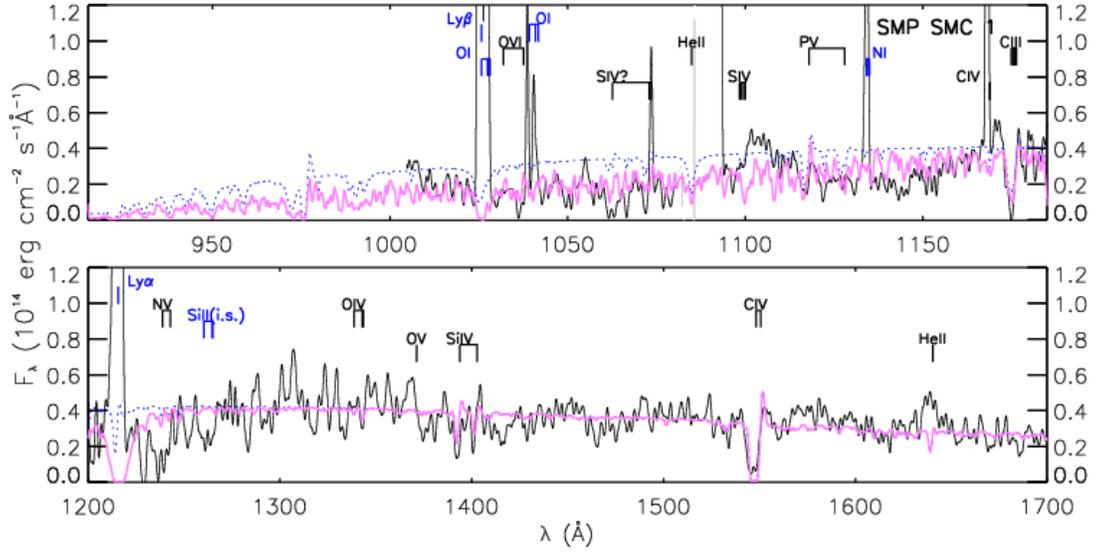}}
%\end{turn}
\caption{\smco: Description follows that of Fig.~\ref{fig:lmc76}.
}
\label{fig:smc1}
\end{center}
\end{figure}

\noindent \emph{\smco\ (Fig.~\ref{fig:smc1})}

Due to uncertainties with the FUSE observations, we have determined
the characteristics of this object by mainly relying on wind features
in its UV spectrum, using \siiv, and \civ\ as diagnostics.  We have
also used the far-UV \ciiib\ feature for guidance.  These features
indicate a relatively low terminal wind velocity of $\vinf \simeq
500$~\kms.  Because of the limited diagnostics, we have assumed
normal, ``H-rich'' SMC abundances.  Models of $\Teff \simeq 37$~kK
produce acceptable fits to the UV spectrum (cooler temperatures weaken
\civ\ unacceptably, and hotter temperatures prohibit the \siiv\ and
\ciiib\ features).  
However, we are unable to attain acceptable fits of
the FUSE spectrum with our stellar models, even when incorporating
various hydrogen absorption models.  The limited wavelength coverage
prevent a determination of \EBMV\ from the spectral fitting, so we
have simply adopted the value of $\EBMV = 0.16$ (from
\citealp{meatheringham:91a}, see Table~\ref{tab:neb}).
We note that in Fig.~\ref{fig:smc1} what appears
to be strong stellar \HeII\ emission feature at \singlet 1640 is
actually multiple narrow emission features (one of which being nebular
\HeII) blended together as a result of the binning for clarity.

\begin{figure}[htbp]
\begin{center}
\epsscale{1.0}
%\begin{turn}{90}
\rotatebox{0}{
\plotone{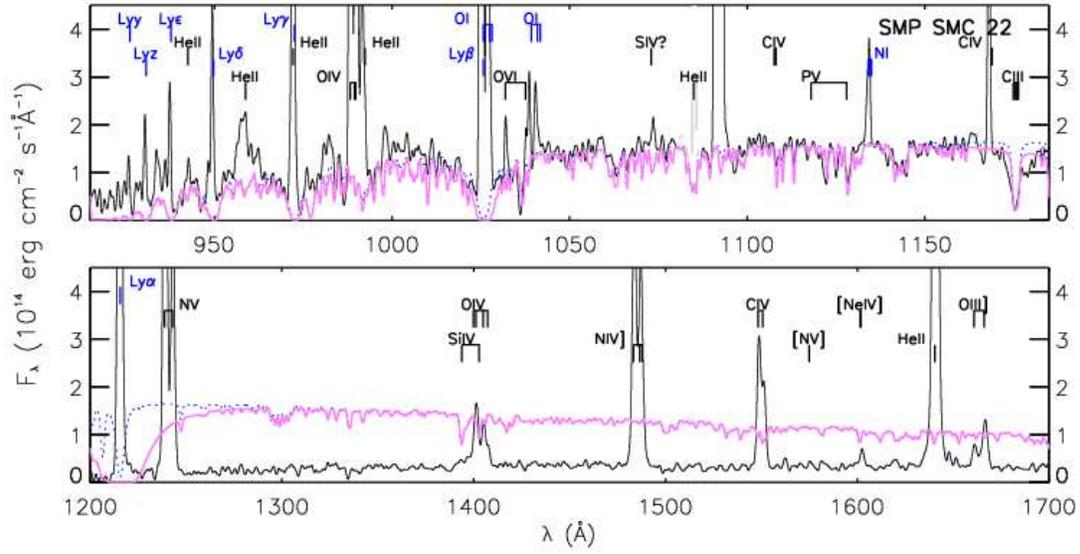}}
%\end{turn}
\caption{\smct: Description follows that of
  Fig.~\ref{fig:lmc76}.  Note: this observation was apparently
  contaminated by a B-star in the aperture, and so does not represent
  the true spectrum of \smct\ (this is the reason for the flux
  discrepancy between the far-UV and UV spectra).
}
\label{fig:smc22}
\end{center}
\end{figure}

\noindent \emph{\smct\ (Fig.~\ref{fig:smc22})}

This star has a very high photoionization temperature of $\Teff =
200$~kK (Table~\ref{tab:shock}).  However, we are able to fit the FUSE
spectrum well with a SMC-metallicity TLUSTY model of $\Teff = 25$~kK
($\logg=3.7$).  Most of the detailed structure is from \FeIII, which
is present in the atmosphere for $\Teff \lesssim 30$~kK With
$z=0.1$~\Zsun, the \ciiib\ feature is a bit weak.  Using LMC
metallicity ($z=0.4$~\Zsun) the line is better fit, but a higher
temperature ($\Teff = 25$) is required to match the \FeIII\ spectrum.
We are able to fit its FUSE spectrum without invoking a hot
circumstellar molecular gas as found in other MC CSPN.

The photometric SMC catalog of \citet{zaritsky:02} lists an object
17.4 arcseconds away from our target (coordinates: 0 58 36.77 -71 36
6.23) with B-V = -0.26, U-B = -0.77.  These colors are consistent with
those of an early B-type star, which would have $\Teff \sim 20$~kK.  A
main sequence star of this subtype would have $R\simeq4$~\Rsun\ and
$\logg\simeq4$.  Using a TLUSTY model with these parameters, and
scaling the flux to the distance of the SMC (assuming a foreground
reddening of 0.05) results in a far-UV flux level of $\simeq
1\E{-14}$~\flam, about 80\% of the flux levels of our FUSE
observations.  Scaling its flux by a factor of $\simeq1.5$ to match
the U-magnitude from the catalog (U$=16.38$) results in essentially
the far-UV flux levels of the FUSE observations.  Given that the FUSE
flux levels are several times higher than the FOS fluxes, we believe
that the most likely explanation is that, due to the significant
target drift experienced during the observation of this target
(\S~\ref{sec:fuvobs}), the light from the B-type star entered the FUSE
aperture to some extent, and the FUSE spectra are dominated by this
other object, rather than the target. Thus we exclude this object from
subsequent analysis.

\section{DISCUSSION}\label{sec:discussion}

\begin{figure}[htbp]
\begin{center}
\epsscale{1.0}
\rotatebox{0}{
\plotone{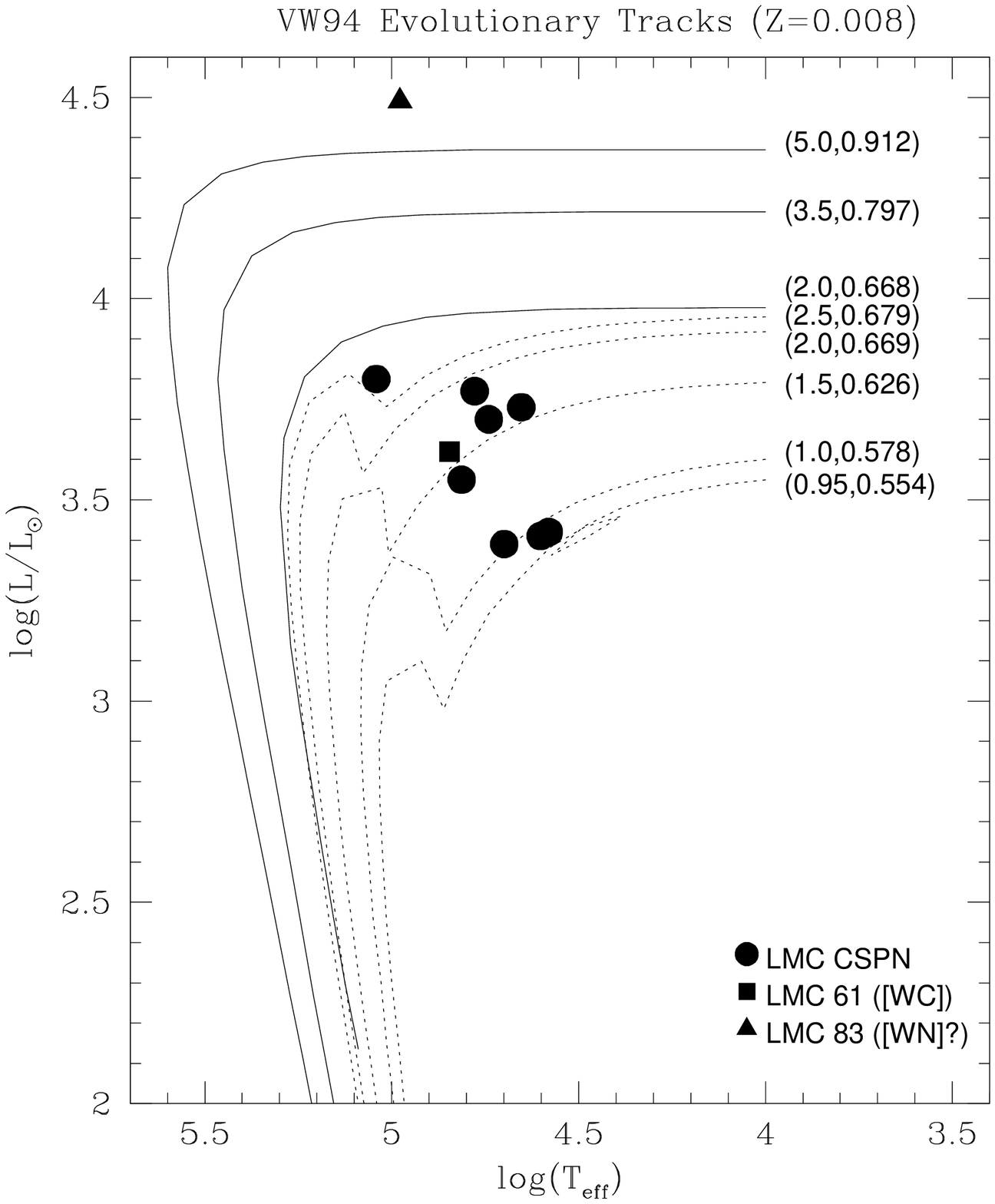}}
\caption{HR diagram of our LMC sample and that of HB04.  Evolutionary
  tracks (Z=0.008) of VW94 are shown, labeled with their initial and
  final masses, with the
  H and He-burning tracks denoted by the solid and dotted lines,
  respectively.  Note our adopted model metallicity corresponds to Z=0.006.
}
\label{fig:tracks_lmc}
\end{center}
\end{figure}

\begin{figure}[htbp]
\begin{center}
\epsscale{1.0}
\rotatebox{0}{
\plotone{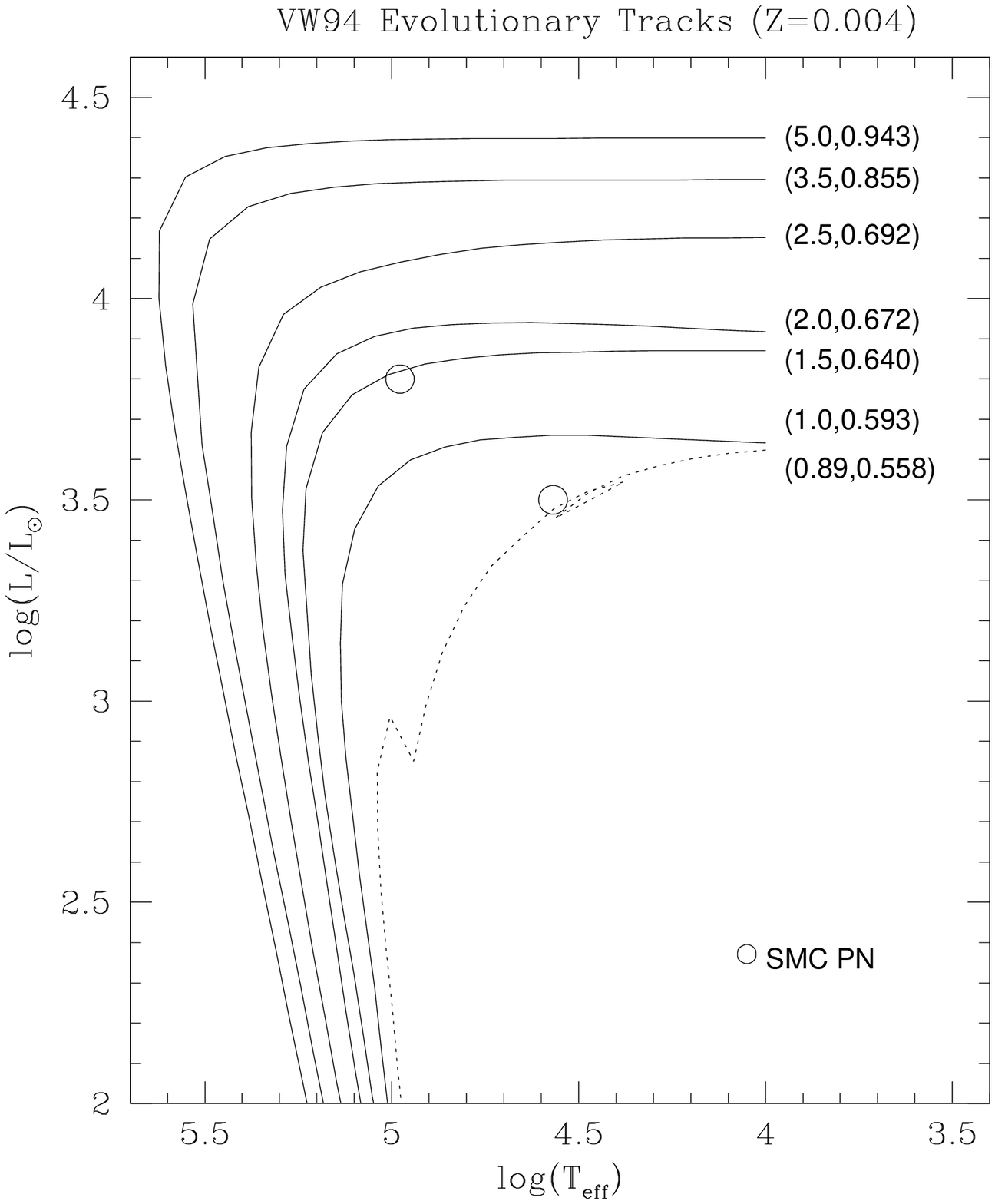}}
\caption{HR diagram of our SMC sample.  Evolutionary
  tracks (Z=0.004) of VW94 are shown, labeled with their initial and
  final masses, with the
  H and He-burning tracks denoted by the solid and dotted lines,
  respectively.  Note our adopted model metallicity corresponds to Z=0.0016.
}
\label{fig:tracks_smc}
\end{center}
\end{figure}

In Table~\ref{tab:shock}, we present a combined summary of our
findings here with those of HB04.  We list temperatures derived from
photo-ionization models
\citep{dopita:91a,dopita:91b,dopita:97,vassiliadis:98} as well as
those from our stellar atmospheres analysis.  We also list the
measured terminal velocity of the wind, and the
morphology of the surrounding PN.  
In Figs.~\ref{fig:tracks_lmc} and \ref{fig:tracks_smc} we plot our
sample along with that of HB04 with the evolutionary tracks of VW94
for LMC and SMC metallicities, respectively (both hydrogen and helium
burning tracks are shown).  Most of the the LMC sample is seen to lie
in the He-burning regime.  VW94 found similar results when they
plotted their tracks with the CSPN samples of
\citet{dopita:91a,dopita:91b} (who derived parameters based on
photoionization models), and attributed this result to the different
timescales of the H and He-burning tracks.  He-burners are expected to
take several thousands of years to migrate across the interval
$\log{L} \sim 4.0$---$2.5$~\Lsun, whereas the H-burners only take $\lesssim
1000$ years to span the same range.  The tracks of VW94 apparently
indicate that, as a consequence of their faster evolution to smaller
luminosities, the LMC H-burners have already faded below the threshold
of FUSE, as our sample represents the expected UV-brightest CSPN.

We found evidence of hot molecular hydrogen in the circumstellar
environments of our previous sample of 7 LMC stars (HB04).  Here we
have made similar findings for our targets.  This is further
indication that hot \Htwo\ in the circumstellar environments of PN is
not uncommon (\eg, see \citealp{herald:02,herald:04b} for Galactic
examples).  \Htwo\ may exist in clumps, shielded from the intense UV
radiation fields by neutral and ionized hydrogen, as appears to be the
case in the Helix nebula \citep{speck:02}.  \citet{speck:02} suggest
that these clumps may form after the onset of the PN phase, arising
from Rayleigh-Taylor instabilities at either the ionization front or
the fast wind shock front.  Theoretical models by \citet{aleman:04}
show that a significant amount of \Htwo\ can survive within the
ionized region of PN, and that the ratio of total \Htwo\ to hydrogen
mass inside the ionized region increases with the temperature of the
central star.  A correlation between the presence of \Htwo\ line
emission and the morphology of the PN has been noted by
\citet{zuckerman:88, kastner:96} (known as ``Gatley's rule'').

For the stars analyzed in this paper, our derived effective
temperatures are mostly in good agreement with those derived from
photoionization models.  For \smce, which shows sparse spectral
diagnostics, a stellar model with a temperature of about the
photoionization temperature ($\sim95$~kK) fits the spectra adequately.
The temperature derived through spectral modeling (by HPGR) of the
enigmatic \lmca\ is significantly lower than through photoionization
models ($\sim100$~kK vs. 170~kK).

\section{CONCLUSIONS}\label{sec:conclusions}

HB04 analyzed 6 LMC CSPN, and here we have analyzed FUSE observations
of 4 additional LMC and 2 SMC CSPN.  We determined the stellar
parameters using the stellar atmosphere code CMFGEN to analyze their
FUSE far-UV and HST UV spectra.  In some cases, we also modeled the
nebular continua (if it contaminates the UV spectra) and the atomic
and molecular hydrogen absorption along the sight-line (which severely
affects the far-UV flux).

To summarize, we find:

\begin{itemize}

\item For those objects with photoionization temperatures between $30
  \lesssim T \lesssim 100$~kK, the temperatures determined via stellar
  modeling are in rough agreement.  For \lmca, our derived effective
  temperature of $\Teff \simeq 95$~kK, is significantly below the
  photoionization temperature of $T=170$~kK.  However, this unique
  object has shown substantial changes in its spectra (and,
  consequently, its parameters) over the past two decades.

\item Most objects often have absorption features in the far-UV
  which could be attributed to very hot ($T \gtrsim2000$~K) \Htwo\ in
  their circumstellar environment.  These temperatures may be due to
  the proximity of the nebulae to the star (see theoretical work of
  \citealp{aleman:04}), or perhaps due to shocks (as suggested by
  \citealp{speck:02}).

\item The majority of our objects have $\vinf \leq 1000$~\kms.  Two of
   the exceptions show signs of chemically enriched winds (\lmca\ \&
   \lmcs).  For the other (\smce) the poor data quality prohibit a
   definitive statement about wind abundances from this spectral
   analysis.

\item We find indications of X-rays in the winds of \lmcs.  We are
  able to attain good fits consistently for the stronger wind
  features, except for the far-UV \ovi\ doublet, which is too weak in
  our models unless X-rays are included in the calculations.  Although
  it has been shown that X-rays are needed to replicate this feature
  for O and early B-stars \citep{bianchi:02,garcia:04,bianchi:06iau2}, this
  is the first case of a similar phenomena in CSPN winds.

\item These FUSE observations add two cases (\lmca\ and \smct) to the
  small number of CSPN displaying nebular \ovi\ (the
  first case, LMC-SMP~62, was reported by HB04).

\end{itemize}

By virtue of their membership of the Magellanic Clouds, the
uncertainties in their distances is small, which carries over to the
derived stellar parameters.  This is a great advantage over Galactic
CSPN, where the distance is the largest source of uncertainty in the
analysis.

Our FUSE observations have allowed us to derive a set of stellar and
wind parameters for CSPN of the Magellanic Clouds that are unhampered by the
distance uncertainties that plague equivalent Galactic studies.  In
addition to revealing the flux of the hot star (which is obscured at
longer wavelengths by the nebular flux), the far-UV observations also
indicate that many of these young CSPN have hot molecular
hydrogen in their nebulae, and are thus also giving insight into their
circumstellar environments and histories.

\acknowledgements

We are grateful to the referee, Klaus Werner, for a careful reading of
the manuscript and for his constructive comments.
We thank John Hillier his help with the CMFGEN code, and Thierry Lanz
and Ivan Hubeny for their help with the TLUSTY code.  We thank Stephan
McCandliss for making his \Htwo\ molecular data available.  We
acknowledge that without the efforts to calculate atomic data of the
OPACITY project, this work would not have been possible.  The SIMBAD
database was used for literature searches.  This work has been funded
by NASA grants NAG-12294 (FUSE-C056) and NAG-13679 (FUSE-D034).
The HST data presented in this paper were obtained
from the Multimission Archive at the Space Telescope Science Institute
(MAST). STScI is operated by the Association of Universities for
Research in Astronomy, Inc., under NASA contract NAS5-26555.

\appendix

\section{APPENDIX: MODEL ATOMS}\label{sec:atomic}

For our \tlusty\ models, the atomic data used come from TOPBASE, the
data-base of the Opacity Project \citep{cunto:93}.   Hydrogen and
Helium were treated in NLTE, and the other included species (C, N, O, Ne,
Na, Mg, Si, S, Ar, Ca, and Fe)  were allowed to contribute
to the total number of particles and charge but their opacity
contribution was neglected in the model atmosphere calculation.

Ions and the number of levels and superlevels included in the \cmfgen\
model calculations are listed in able~\ref{tab:ion_tab}.  The atomic data
come from a variety of sources, with the Opacity Project
\citep{seaton:87,opacity:95,opacity:97}, the Iron Project
\citep{pradhan:96,hummer:93}, \citet{kurucz:95}\footnote{See
http://cfa-www.harvard.edu/amdata/ampdata/amdata.shtml} and the Atomic
Spectra Database at NIST Physical Laboratory being the principal
sources.  Much of the Kurucz atomic data were obtained directly from
CfA \citep{kurucz:88,kurucz:02}.  Individual sources of atomic data
include the following: \citet{zhang:97}, \citet{bautista:97},
\citet{becker:95b}, \citet{butler:93}, \citet{fuhr:88},
\citet{luo:89a}, \citet{luo:89b}, Mendoza (1983, 1995, private
communication), \citet{mendoza:95},
\citet{nussbaumer:83,nussbaumer:84}, \citet{peach:88}, Storey (1988,
private communication), \citet{tully:90}, and
\citet{wiese:66,wiese:69}.  Unpublished data taken from the Opacity
Project include: \FeVI\ data (Butler, K.) \FeVIII\ data (Saraph and
Storey) and C. Mendoza (\FeIX, \FeX).

%-----------------------------------------------------------------------

%%\newpage

%---------------------figures-------------------------------------------

\clearpage

\clearpage

\clearpage

\clearpage

\clearpage

\clearpage

%\clearpage

\clearpage

%\clearpage

\clearpage

%\clearpage

\clearpage

%\clearpage

\clearpage

\clearpage

\clearpage

\clearpage

%---------------------tables--------------------------------------------
%\newpage

\begin{deluxetable}{lcclccccc}
%\rotate
\tabletypesize{\scriptsize}
\tablecolumns{10}
%\tablewidth{0pc}
\tablecaption{FUSE C056 and D034 program stars, and FUSE detector segments utilized\label{tab:fuseobs}}
\tablehead{
\colhead{Name} & \colhead{R.A.} & \colhead{Dec} & 
\colhead{Archive} & \colhead{Date(s)} &
\multicolumn{4}{c}{Wavelength Range (\AA)} \\
\colhead{} &\colhead{(2000)}&\colhead{(2000)}& 
\colhead{Name}    &\colhead{}      &  
\colhead{SiC1}  &\colhead{SiC2} &\colhead{LiF1} &\colhead{LiF2}
}
\startdata
\lmco  & 04 38 34.7 & -70 36 43.9 & D0340101 & 7/4/03 & 
920--993 & 916--1000 & 1000-1081,1095--1136 & 1087-1183,1020-1075 \\
\lmca & 05 36 20.80 & -67 18 08.0 & D0340201 & 7/6/03 &
915--993 & 915--1006 & 1000--1084,1093--1184 & 1020--1075,1088--1183 \\
\lmcs & 05 33 56.20 & -67 53 08.4 & D0340301 &4/29/03 &
915-993 & 915-1013 & 1005-1091,1087--1196 & 1005--1082,1079--1186 \\
           &          &           & D0340302 & 4/30/03& 
`` & `` & `` & `` \\
\lmcf & 05 20 51.82 & -67 05 43.1 & D0340401 &  7/3/03 &
915--993 & 916--1006 & 1000--1082,1094--1136 & 1028--1082,1086--1183 \\
\lmct & 05 08 03.47 & -68 40 16.5 & D0340501 & 7/8/03& 
- & - & 1005--1082,1086--1183 & - \\
\smco  & 00 23 58.80 & -73 38 04.1  & C0560101 & 6/11/02 &
- & - & 1005-1084,1093-1187 & - \\
\smce  & 00 34 21.90 & -73 13 20.8  & C0560201 & 6/12/02 &
915-993 & 915-1005 & 1105-1184,1093-1155 & 1005-1082, 1088-1183 \\
\smcf  & 00 41 21.70 & -72 45 18.1  & C0560402 & 8/5/02  \\
           &             &              & C0560401 & 8/3/02  \\
\smct & 00 58 37.22 & -71 35 48.8  & C0560301 & 7/25/02 & 
915-992 & - & 992-1084,1093-1187 & - \\
\enddata
%\tablecomments{}
%\tablenotetext{a}{ ``d'' denotes calibration with default pipeline
%  background subtraction, ``s'' with scattered light background model
%  (see text).}
\end{deluxetable}

%\begin{table}
\begin{deluxetable}{lccccc}
\tabletypesize{\scriptsize}
\tablecolumns{6}
%\tablewidth{0pc}
\tablecaption{Archive UV data sets (FOS and STIS)\label{tab:uvobs}}
%\caption{Archive UV data sets (FOS and STIS)}\label{tab:uvobs}
%\begin{tabular}{lccccc}
%\hline
\tablehead{
\colhead{Name} & \colhead{Instrument} & \colhead{Dataset} &
\colhead{Date} & \colhead{Grating} & \colhead{$\lambda$ (\AA)}
}
%       &            &   &      &         & (\\
%\hline
\startdata
\lmcs & FOS & Y1C10B03T & 9/22/93 & G130H & 1182-1600\\
           & FOS & Y1C10B03T & 9/22/93 & G190H & 1600-2330\\
\lmca & STIS &  O5IM01010 & 02/04/00 & G140L & 1182-1720 \\
       & STIS & O5IM01020 & 02/04/00 & G230L & 1720-3150 \\
       & STIS & O5IM02010 & 02/04/00 & G430L & 3150-5200 \\
       & STIS & O5IM02010 & 02/04/00 & G430L & 3150-5200 \\
       & STIS & O55R01010 & 01/08/99 & G140L & 1182-1720 \\
       & STIS & O55R01020 & 01/08/99 & G230L & 1720-3150 \\
       & STIS & O55R01030 & 01/08/99 & G430L & 3150-5200 \\
       & STIS & O55R02020 & 01/08/99 & G750L & 5265-10260\\
\lmco & FOS & Y3EV0106P & 11/15/96 & G130H & 1182-1600\\
          & FOS & Y3EV0103P & 11/15/96 & G190H & 1600-2300\\
          & FOS & Y3EV0104P & 11/15/96 & G270H & 2300-3280 \\      
          & FOS & Y3EV0105P & 11/15/96 & G400H & 3280-4822 \\      
\lmcf & FOS & Y3EV0805T & 11/10/96 & G130H & 1182-1600\\
           & FOS & Y3EV0802T & 11/10/96 & G190H & 1600-2300\\
           & FOS & Y3EV0803T & 11/10/96 & G270H & 2300-3280 \\      
           & FOS & Y3EV0804T & 11/10/96 & G400H & 3280-4822 \\      
\lmct & FOS & Y2Y00204T & 10/22/95 & G130H & 1182-1600\\
           & FOS & Y2Y00205T & 10/22/95 & G190H & 1600-2300\\
           & FOS & Y2Y00206T & 10/22/95 & G270H & 2300-3280 \\      
           & FOS & Y2Y00207T & 10/22/95 & G400H & 3280-4822 \\      
\hline
\smco & FOS & Y1C10103T & 5/3/93 & G130H & 1182-1600\\
          & FOS & Y1C10104T & 5/3/93 & G190H & 1600-2330\\
\smce & FOS & Y1C10203T & 11/26/93 & G130H & 1182-1600\\
          & FOS & Y1C10204T & 11/26/93 & G190H & 1600-2330\\
\smcf & FOS & Y2Y00904T & 6/12/96 & G130H & 1182-1600\\
          & FOS & Y2Y00905T & 6/12/96 & G190H & 1600-2330\\
          & FOS & Y2Y00906T & 6/12/96 & G270H & 2300-3280 \\      
          & FOS & Y2Y00907T & 6/12/96 & G400H & 3280-4822 \\      
\smct & FOS & Y2N30104T & 3/20/95 & G130H & 1182-1600\\
          & FOS & Y2N30105T & 3/20/95 & G190H & 1600-2330
%\hline
\enddata
%\end{tabular}
%\end{table}
\end{deluxetable}
\begin{deluxetable}{lccccccccccc}
%\rotate
\tabletypesize{\scriptsize}
%\tabletypesize{\footnotesize}
\tablecolumns{12}
%\tablewidth{0pc}
\tablecaption{Nebular Parameters \label{tab:neb}}
\tablehead{
\colhead{Name} & 
\colhead{$\theta^{aa}$} &
\colhead{$r_{neb}^{bb}$} &
\colhead{$v_{exp}(\rm{OIII/OII})^{a}$} &
\colhead{$\tau_{dyn}^{cc}$} &
\colhead{$n_{e}([\rm{OII/SII}])^{b}$} & 
\colhead{\Telec(O$^{+2}$/N$^{+1}$)$^{b}$} & 
\colhead{$\log(F_{H\beta})^{h}$} &
\colhead{\EBMV(c$_{H\beta}$)$^{b}$} & 
\colhead{He/H$^c$} & 
\colhead{He$^{2+}$/He$^+$} &
\colhead{$v_{LSR}$} \\
\colhead{} & 
\colhead{(\arcsec)} & 
\colhead{(pc)} & 
\colhead{(\kms)} & 
\colhead{(kyr)} &
\colhead{(10$^3$ cm$^{-3}$)} & 
\colhead{(kK)} & 
\colhead{()} & 
\colhead{(mag)} & 
\colhead{} & 
\colhead{} &
\colhead{(\kms)}
}
\startdata
\lmco  &0.33$^{s}$& 0.04& 17.2/23.1 & 2.3 & 3.44/3.32 & 11.5/-&
-12.46 (\textbf{-12.5})& 0.11& 0.146& -$^e$ (\textbf{0.0}) & 209.1$^a$ \\
\lmct & 0.68 & 0.083$^{d}$ & 35.9/49.3 & 2.3 &5.46/4.06 &20.0/- &
-12.71 & 0.05 & 0.160 & 1.53$^e$& 228.2$^a$\\
\lmcf &0.64$^{s}$ & 0.08& 35.0/55.7 & 2.2 & 5.37/- & 11.2/- &
-12.71 (\textbf{-12.81})&0.04 & 0.120& 0.23$^e$ & 284.1$^a$\\
\lmcs & $<$0.16& $<$0.02 & 29.0$^{xx}$/32.9 & 0.7&13.7/- &12.0/- &
-12.54&0.22 & 0.075 & -$^e$ (\textbf{0.0})& 262.8$^a$\\
\lmca & 2.34 & 0.292$^{d,x}$ & 82.9$^{xx}$/80.9 &3.4 &2.46/2.22
&17.4/11.7 & -12.65&0.10& 0.137 & 1.15$^e$ & 274.5/300$^k$\\
\hline
\smco  & 0.31 & 0.045$^{d}$ & 15.4/16.1 & 2.9& 9.59/- & 10.7/- &
-12.77 (\textbf{-12.89})& 0.16 &
0.080 & -$^e$/0.155$^{g}$  & 138.0$^j$ \\
\smce  & 0.49 & 0.072$^{d}$ & 32.9/-    & 2.1& 1.96/- & 13.5/- &
-13.06 (\textbf{-12.78}) & 0.03 &
0.160 & 0.07$^e$/1.66$^g$& 98.2$^j$\\
\smcf  & 0.42& 0.061$^{d}$ & 29.2/43.6 & 2.0& 3.7/3.34 & 14.5/- & -12.81& 0.07 &
0.130 & 0.60$^{e}$& 101.2$^j$\\
\smct  & 0.52& 0.075$^{d}$ & 50.9/51.2 & 1.5& 5.51/2.33 &
26.6/11.6 & -12.88 & 0.11&0.135$^{g}$& & 144.2$^j$ \\
\enddata
\tablenotetext{aa}{Calculated from $r_{neb}$  using LMC/SMC distances of 50.6/60.0~kpc}
\tablenotetext{bb}{From their ``D(edge)''}
\tablenotetext{cc}{$\tau_{dyn}=r_{neb}/v_{exp}$}
\tablenotetext{dd}{Line fit with two components}
\tablenotetext{ee}{From [\SII] \citep{meatheringham:91b}}
\tablenotetext{xx}{Multiple components}
\tablenotetext{xxx}{Avg. of 3 components}
\tablerefs{
(a): \citet{dopita:88}
(b): \citet{meatheringham:91a,meatheringham:91b}
(c): \citet{dopita:91a,dopita:91b}
(d): \citet{vassiliadis:98}
(e): \citet{monk:88}
(f): \citet{dopita:96}
(g): \citet{henry:89}
(h): \citet{meatheringham:88b}
(j): \citet{dopita:85}
(k): \citet{pena:04}
(s): \citet{shaw:01}
(x): approx of three vals given
}
\end{deluxetable}

\begin{deluxetable}{cccccccccc}
\tabletypesize{\scriptsize}
\tablecolumns{10}
\tablecaption{Abundances of grid models\label{tab:abund}}
\tablehead{
\colhead{Grid}    & \colhead{\XH} & \colhead{\XHe} & \colhead{\XC} & \colhead{\XN} & \colhead{\XO} & \colhead{\XS}i & \colhead{\XP} & \colhead{\XS} & \colhead{\XFe} 
}
\startdata
LMC H-deficient & - & 0.54 & 0.37 & 0.01 & 0.08 & 2.80\E{-4} & 2.45\E{-6} & 1.53\E{-4} & 5.44\E{-4}\\
LMC H-rich & 0.71 & 0.28 & 1.22\E{-3} & 4.40\E{-4} & 3.82\E{-3} &
2.80\E{-4} & 2.45\E{-6} & 1.53\E{-4} & 5.44\E{-4}\\
SMC H-rich & 0.71 & 0.28 & 3.05\E{-4} & 1.10\E{-4} & 9.55\E{-4} &
7.00\E{-5} & 6.12\E{-7} & 3.65\E{-5} & 1.36\E{-4}
\enddata
\end{deluxetable}

\begin{deluxetable}{cccccccccc}
\tabletypesize{\scriptsize}
\tablecolumns{10}
%\tablewidth{0pc}
\tablecaption{\HI\ and \Htwo\ parameters{\label{tab:hydrogen}}}
\tablehead{
\colhead{} & \multicolumn{2}{c}{$\HI^{IS + circ}$} & \colhead{} & \multicolumn{2}{c}{$\Htwo^{IS}$} &
\colhead{} & \multicolumn{2}{c}{$\Htwo^{circ}$} & \colhead{}\\
\cline{2-3}\cline{5-6}\cline{8-9}
\colhead{Star} & \colhead{\logN} & \colhead{$T$} & \colhead{} &
\colhead{\logN} & \colhead{$T$} & \colhead{} & \colhead{\logN} &
\colhead{$T$} & \colhead{\EBMV (\HI)}\\
\colhead{} &\colhead{(cm$^{-2}$)} &\colhead{(K)} & \colhead{} &
\colhead{(cm$^{-2}$)} &\colhead{(K)} & \colhead{} & \colhead{(cm$^{-2}$)}
&\colhead{(K)} & \colhead{(mag)} \\
}
\startdata
\lmcs & $21.00^{+0.3}_{-0.3}$ & 80 && $18.70^{+0.3}_{-0.3}$ &
80 &&$16.70^{+0.0}_{-0.0}$  & $2000\pm1000$ & 0.2 \\
\lmco & $20.70^{+0.0}_{-0.0}$ & 80 && $18.00\pm1.0$ &
80 &&$17.00^{+0.4}_{-0.3}$  & $3000\pm1500$ &  0.1\\
\lmcf & $20.40^{+0.3}_{-0.4}$ & 80 && $19.7^{+0.3}_{-0.3}$ &
80 &&$16.70^{+0.7}_{-0.7}$  & $3000\pm1500$ &  0.05\\
\lmca & \textbf{$20.80^{+0.2}_{-0.4}$} & 80 && $16.70^{+0.7}_{-0.7}$ &
80 &&$16.70^{+0.3}_{-0.4}$  & $2000\pm1000$ & 0.04 \\
\lmct & - & - && - & -  &&-  & - & - \\
\hline
\smco & 21.9 & 80 && - & - &&-  &- & -\\
\smce  & $21.7$ & 80 && $19.40^{+0.3}_{-0.4}$ 2.1&
80 &&$16.70^{+0.3}_{-0.4}$  & $2000\pm1000$ &  1.0\\
\smct & $20.00^{+0.3}_{-0.3}$ & 80 && $16.70^{+0.3}_{-0.4}$ &
80 &&-  & - &  0.02\\
\enddata
\end{deluxetable}

\begin{deluxetable}{lcccccccc}
%\rotate
\tabletypesize{\footnotesize}
\tablecolumns{11}
%\tablewidth{0pc}
\tablecaption{Derived Stellar Parameters \label{tab:mod_param}}
\tablehead{
\colhead{Name} & 
\colhead{Model} & 
\colhead{\Teff} & 
\colhead{$\log{L}$} & 
\colhead{\Rstar} & 
\colhead{$\log{\Mdot}$} & 
\colhead{\vinf} &
\colhead{\logg} & 
\colhead{\EBMV} \\
\colhead{} & 
\colhead{Abundance} & 
\colhead{(kK)} & 
\colhead{(\Lsun)}& 
\colhead{(\Rsun)} & 
\colhead{(\Msunyr)} & 
\colhead{(\kms)} & 
\colhead{(cgs)}& 
\colhead{(mag)}  
}
\startdata
\lmca & LMC~83 & 95 & $4.49^{+0.17}_{-0.21}$ & $0.47\pm0.1$ & $-5.71^{+0.13}_{-0.16}$ & 1600  & 6.0 &  0.08\\
\lmco & H-poor & $65^{+5}_{-10}$ & $3.61^{+0.08}_{-0.20}$
&$0.70^{+0.10}_{-0.05}$ & $-7.52^{+0.26}_{-0.21}$ & $800\pm100$  & 5.0  & 0.13\\
\lmco & H-rich & $65^{+5}_{-10}$ & $3.46^{+0.08}_{-0.20}$
&$0.66^{+0.10}_{-0.04}$ & $-7.46^{+0.26}_{-0.21}$ & $800\pm100$  & 5.0 & 0.13\\
\lmcs & H-poor & $50\pm5$ & $3.39^{+0.29}_{-0.22}$ &$0.67\pm0.1$ &
$-7.49^{+0.37}_{-0.33}$ & $1250\pm250$ & 4.7 & 0.03\\
\lmcf & H-rich & $110\pm10$ & $3.80^{+0.26}_{-0.34}$ &$0.22^{+0.03}_{-0.04}$ & $-8.2^{+0.30}_{-0.20}$ & $500\pm100$  & 6.0 & 0.04\\

\lmct & - & - & - &- & - & - & -  & 0.05\\
\hline
\smco & H-rich & 37$^{+3}_{-2}$ & 3.5$^{+0.3}_{-0.2}$ & $1.3\pm0.2$ &-7.31$\pm0.10$ & $500\pm100$ & 4.9  &  0.16\\
\smce & H-rich & 95$\pm$5 & 3.8$^{+0.3}_{-0.2}$ &
$0.32\pm0.05$ & $-7.8\pm0.1$ & $\ge2000$ & 5.7 &  0.03\\
\enddata
%\tablenotetext{a}{From evolutionary tracks of \citet{vassiliadis:94}.}
\end{deluxetable}
\begin{deluxetable}{lcccc}
%\rotate
%\tabletypesize{\scriptsize}
\tabletypesize{\footnotesize}
\tablecolumns{12}
%\tablewidth{0pc}
\tablecaption{Summary Table \label{tab:shock}}
\tablehead{
\colhead{Name} & 
\colhead{$\Teff^{aa}$} &
\colhead{$T_{phot}^{bb}$} &
\colhead{$\vinf^{aa}$} & 
\colhead{Morphology} \\
\colhead{} &  
\colhead{(kK)} & 
\colhead{(kK)} & 
\colhead{(\kms)} &  
\colhead{} \\
}
\startdata
LMC-SMP~2 & 38 & 39 & 700  & Spherical$^c$, Round$^a$\\
LMC-SMP~85 & 40 & 45 & 700  & Round$^a$\\
LMC-SMP~62 & 45 & 127 & 1000 & Bipolar w/central condensation$^b$\\
\lmcs & 50 & 50 & 1250 & Spherical$^c$,  Round$^a$\\
LMC-SMP~67 & 55 & 46 & 1000 & Bipolar w/central condensation$^b$\\
LMC-SMP~23 & 60 & 65 & 1100 & Ring$^b$, Round$^a$\\
\lmco  & 65 & 66 & 800 & \\
LMC-SMP~61 & 70 & 59 & 1300 & Spherical$^b$ \\
\lmcf & 110 & 100 & 500  & Elliptical (bipolar core?)$^d$\\
\lmca & 95 & 170 & 1600 & Spherical$^b$, Bipolar Ring$^b$\\
\lmct & - & 200 & - &  \\
\hline
\smco & 37 & 41  & 500 &  Spherical$^{b,c}$, Round$^a$\\
\smce & 95 & 92  & $>2000$  & Bipolar Ring$^{b,c}$, Bipolar w/bipolar core$^a$\\
\smct & - & 200 & - &  Bipolar Ring$^b$, Bipolar w/bipolar core$^a$\\
\enddata
\tablerefs{
(aa): This paper.
(bb): \citet{dopita:91a,dopita:91b,dopita:97,vassiliadis:96s,vassiliadis:98s}
(a): \citet{stanghellini:99}
(b): \citet{vassiliadis:98}
(c): \citet{dopita:96}
(d): \citet{shaw:01}
}
\end{deluxetable}

\begin{deluxetable}{lccccccccccc}
\tabletypesize{\footnotesize}
\tablecolumns{12}
%\tablewidth{0pc}
\tablecaption{Levels and superlevels for model ions\label{tab:ion_tab}}
\tablehead{
\colhead{Elemen}t & \colhead{I} & \colhead{II} & \colhead{III} &
\colhead{IV} & \colhead{V} & \colhead{VI} & \colhead{VII} &
\colhead{VIII} & \colhead{IX} & \colhead{X} & \colhead{XI}
}
\startdata
H  & 20/30 & 1/1 & & & & & & & & \\
He & 40/45 & 22/30 & 1/1\\
C  & & & 30/54 & 13/18 & 1/1\\
N  & & & & 29/53 & 13/21 & 1/1\\
O  & & & & 29/48 & 41/78 & 13/19 & 1/1\\
Si & & & & 22/33 & 1/1\\
P  & & & & 36/178 & 16/62 & 1/1\\
S  & & & & 51/142 & 31/98 & 28/58 & 1/1\\
Fe & & & & 51/294 & 47/191 & 44/433 & 41/254 & 53/324 & 52/490 &43/210 &1/1
\enddata
\end{deluxetable}

\end{document}